   \documentclass[twocolumn,showpacs,preprintnumbers,amsmath,amssymb]{revtex4}
   \usepackage{graphicx,xcolor}    
\usepackage{epstopdf,epsfig}
	\usepackage{bm}
   \usepackage{dcolumn}
   \usepackage{natbib}
	\usepackage{physics}

\begin{document}
   	
   	\title{High energy shift in the optical conductivity spectrum of the bilayer graphene}
   	
   	\author{V. Apinyan\footnote{Corresponding author. Tel.:  +48 71 3954 284; E-mail address: v.apinyan@int.pan.wroc.pl.}, T. K. Kope\'{c}}
   	\affiliation{Institute of Low Temperature and Structure Research, Polish Academy of Sciences\\
   		PO. Box 1410, 50-950 Wroc\l{}aw 2, Poland \\}
   	
   	\date{\today}

\begin{abstract}
%
We calculate theoretically the optical conductivity in the bilayer graphene by considering Kubo-Green-Matsubara formalism. Different regimes of the interlayer coupling parameter have been considered in the paper. We show that the excitonic effects substantially affect the optical conductivity spectrum at the high-frequency regime when considering the full interaction bandwidth, leading to a total suppression of the usual Drude intraband optical transition channels and by creating a new type of optical gap. We discuss the role of the interlayer coupling parameter and the Fermi level on the conductivity spectrum, going far beyond the usual tight-binding approximation scheme for the extrinsic bilayer graphene. 
   	\end{abstract}

   	\maketitle

 \renewcommand\thesection{\arabic{section}}
   	
\section{\label{sec:Section_1} Introduction}
%
The introduction of two-dimensional materials and the possibility to control their optical properties brings the new and novel high valued technological applications in nanophotonics, optoelectronics and solar cells \cite{cite_1}.  
The optical properties of the monolayer (MG) and bilayer graphene (BLG) structures are of great importance in the context of the modern technological applications in the infrared, visible and terahertz range of the frequency spectrum. By applying the external gate voltage one can modify the density of charge carriers and the position of the Fermi level in these systems \cite{cite_2}. The imposition of external electric field tunes the bilayer graphene from the semimetallic to the semiconducting state \cite{cite_3}, by allowing for novel terahertz devices \cite{cite_4} and transistors \cite{cite_5}. On the other hand, doped, or electrically tuned graphene and bilayer graphene allow for the spontaneous chiral symmetry breaking states, reflecting in the form of the gapped states in the fermionic quasiparticle spectrum \cite{cite_6, cite_7, cite_8, cite_9, cite_10, cite_11, cite_12}. The spectacular optical properties of bilayer graphene make it as a promising material for infrared optoelectronics. The optical transitions in the system can be alternated, after electrical gating of graphene and BLG \cite{cite_13}, and the effects are very similar to the case of the charge transport in the field-effect transistor constructions \cite{cite_14, cite_15}.
The optical and charge transport properties in the BLG structures have been widely studied both theoretically \cite{cite_16, cite_17, cite_18, cite_19, cite_20, cite_21, cite_22, cite_23, cite_24, cite_25, cite_26, cite_27, cite_28, cite_29, cite_30} and experimentally \cite{cite_31, cite_32, cite_33, cite_34}. The optical conductivity has been crucial for the optical determination of the band gap formation in the BLG \cite{cite_35, cite_36, cite_37,cite_38}. In the visible range of the spectrum, the band gap was calculated in Ref.\onlinecite{cite_39}. A very comprehensive comparison of different microscopic models for the BLG and also the analysis of the optical transitions in the BLG heterostructures is done in Ref.\onlinecite{cite_40}. In particular, the effects of short ranged scatterers and screened Coulomb-impurities has been discussed and both intrinsic and extrinsic case of bilayer graphene has been considered in details. The existence of the resonant excitons and bound electron-hole (e-h) pairs in graphene and bilayer graphene structures has been confirmed after recent first principle calculations and experimental studies \cite{cite_41,cite_42}. In the main part of those studies, the excitonic effects enhance from broadly resonant excitonic
states, consisting of $\pi$ and $\pi^{\ast}$ bands corresponding the low-frequency regime (up to $10$ eV) and with the extremely short lifetimes. The bound electron-hole (e-h) pairs are of particular importance because of their well-defined binding energies, which decides the efficiency of photovoltaic solar cells \cite{cite_43,cite_44}. 

A new type of high-frequency excitonic effects have been obtained in Refs.\onlinecite{cite_45, cite_46}, in the high-frequency regime (9$\sim$ 20 eV), in the optical spectra of BLG. For the intrinsic graphene, those excitonic effects are the consequence of the unique parallel $\sigma$ and $\pi^{\ast}$ bands, which results in a giant joint density of states. In contrast to ordinary semiconductors or insulators, the excitonic effects in graphene are stronger in the high-energy range of the spectrum \cite{cite_46}. Nevertheless, the most of the theoretical studies on the excitonic effects, discussed above, are limited only to the intrinsic case of the BLG, i.e., when the Fermi level is set to be zero. Therefore, it is of a fundamental interest to consider the excitonic effects in the high-frequency regime at which the BLG shows significant optical activities.

In the present paper, we consider the excitonic effects on the optical conductivity in the BLG and we show how the solution of the Fermi energy in the system affects the optical conductivity spectrum when passing through the balanced charge neutrality point (CNP) in the interacting BLG. We considered the optical conductivity in the BLG by taking into account the intralayer Coulomb interaction and a wide range of the local interlayer Coulomb repulsion between the electrons on different sublattices of BLG. We show how the excitonic effects shift the spectrum of the optical conductivity and absorption spectrum towards the UV region of photonic frequencies. We suppose the half-filling condition in each of the BLG sheets and for all different values of the interlayer coupling parameter. When evaluating the ac-conductivity in the bilayer graphene, we use the numerical values for the excitonic gap parameter and chemical potential, obtained in Refs.\onlinecite{cite_11, cite_12}. We calculate the optical conductivity using the Kubo-Green-Matsubara formalism \cite{cite_47} and we include the local excitonic pairing between the layers of the BLG. We consider different regimes of the interlayer Coulomb interaction parameter, corresponding to different states in the BLG: from semimetallic to semiconducting. 

The structure of the paper is following: in the Section \ref{sec:Section_2}, we introduce the Hubbard model for the BLG. In the Section \ref{sec:Section_3}, we obtain the general expression for the longitudinal optical polarization function and the ac optical conductivity. In the Section \ref{sec:Section_4}, we discuss the numerical results. Finally, in the Section \ref{sec:Section_5}, we give a conclusion to our manuscript. The Appendix, at the end of the paper, is devoted to the calculation of the current-current correlation function. 
%
\section{\label{sec:Section_2} The model Hamiltonian}
%
\begin{figure}
	\begin{center}
		\includegraphics[scale=0.4]{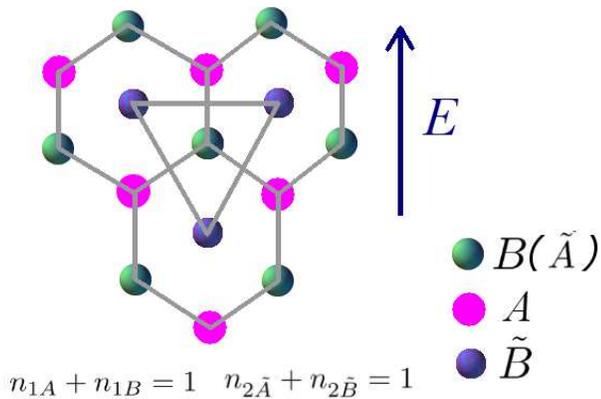}
		\caption{\label{fig:Fig_1}(Color online) The AB-stacked bilayer graphene structure with the applied longitudinal electric field component ${\bf{E}}$. Different sublattice sites positions are shown and the half-filling condition is written at the bottom of the picture, for both layers in the BLG}.
	\end{center}
\end{figure} 
%
The vertical from top-view of the AB-stacked bilayer graphene is presented in Fig.~\ref{fig:Fig_1}. Different lattice site positions in the layer-1 are shown by letters $A$ and $B$ and in the layer-2 by tilde letters $\tilde{A}$ and $\tilde{B}$. 
The bilayer Hubbard Hamiltonian for the presented AB-stacked bilayer graphene structure without the external electric field is given by 
\begin{eqnarray}
H=H_{||}+H_{\perp}+H_{U-V}.
	\label{Equation_1}
\end{eqnarray}
First two terms, in Eq.(\ref{Equation_1}), form the well-known tight binding model. The first term in the Hamiltonian describes the usual hopping of the electrons between the nearest neighbor lattice sites, in a given layer, i.e.,
\begin{eqnarray}
H_{||}&=&-\gamma_0\sum_{\left\langle {\bf{r}}{\bf{r}}'\right\rangle}\sum_{\sigma}\left[{a}^{\dag}_{\sigma}({\bf{r}})b_{\sigma}({\bf{r}}')+h.c.\right]
\nonumber\\
&-&\gamma_0\sum_{\left\langle {\bf{r}}{\bf{r}}'\right\rangle}\sum_{\sigma}\left[{\tilde{a}}^{\dag}_{\sigma}({\bf{r}})\tilde{b}_{\sigma}({\bf{r}}')+h.c.\right]
\nonumber\\
&-&\sum_{{\bf{r}}\sigma}\sum_{\ell=1,2}\mu_{\ell}n_{\ell\sigma}({\bf{r}}).
\label{Equation_2}
\end{eqnarray}
Here, ${a}^{\dag}_{\sigma}({\bf{r}})$ (${a}_{\sigma}({\bf{r}})$) and ${b}^{\dag}_{\sigma}({\bf{r}})$ (${b}_{\sigma}({\bf{r}})$) are the electron creation (annihilation) operators and the same operators with the tilde notations refer to the layer-2 in BLG. The parameter $\gamma_0$ describes the intralayer hopping in the graphene sheets (the most realistic value of it is $\gamma_0=3$ eV, given in Ref.\onlinecite{cite_31}. The chemical potential term in Eq.(\ref{Equation_2}), has been added in order to deal with the Grand canonical ensemble. Initially, we postulate the equilibrium state in the BLG, i.e., $\mu_1=\mu_2\equiv \mu$. The density operator $n_{\ell\sigma}({\bf{r}})$, in the last term in Eq.(\ref{Equation_2}), is $n_{\ell\sigma}({\bf{r}})=\sum_{\eta_{\ell}}n_{\eta_{\ell}\sigma}({\bf{r}})$, where we putted $\eta_{\ell}=a,b$ for $\ell=1$ and $\eta_{\ell}=\tilde{a},\tilde{b}$ for $\ell=2$ and $n_{\eta_{\ell}\sigma}({\bf{r}})=\eta^{\dag}_{\ell\sigma}({\bf{r}})\eta_{\ell\sigma}({\bf{r}})$ is the usual density operator for a given sublattice, in the layer $\ell$. The second term $H_{\perp}$ is responsible for the interlayer hopping in the BLG and is given as
	\begin{eqnarray}
	H_{\perp}=&-&\gamma_1\sum_{{\bf{r}}\sigma}\left[{{b}}^{\dag}_{\sigma}({\bf{r}})\tilde{a}_{\sigma}({\bf{r}})+h.c.\right].
	\nonumber\\
	\label{Equation_3}
	\end{eqnarray}
	The parameter $\gamma_1$ describes the interlayer hopping between different layers in BLG. The interaction part in the system is given by the last term in the Hamiltonian in Eq.(\ref{Equation_1}). Namely, we have 
\begin{eqnarray}
H_{U-V}&=&U\sum_{{\bf{r}}}\sum_{\ell\eta}\left[\left(n_{\ell\eta\uparrow}-1/2\right)\left(n_{\ell\eta\downarrow}-1/2\right)-1/4\right]
\nonumber\\
&+&V\sum_{{\bf{r}}\sigma\sigma'}\left[\left(n_{1b\sigma}({\bf{r}})-1/2\right)\left(n_{2\tilde{a}\sigma'}({\bf{r}})-1/2\right)-1/4\right].
\nonumber\\
\label{Equation_4}
\end{eqnarray}
Parameters $U$ and $V$ in Eq.(\ref{Equation_4}) signify local intralayer and interlayer Coulomb interactions in the BLG structure. Indeed, as we will see later on, the consideration of the local interlayer coupling simplifies the problem substantially. This becomes clear after transforming the Hamiltonian in Eq.(\ref{Equation_1}) into the Fourier space representation with appropriate linearisation of the fermionic action of the BLG. We consider the parameter $\gamma_0$ as the unit of the energy scale in the considered problem.
%
\subsection{\label{sec:Section_2_1} The interaction term $U-V$}
%
Here, we will show how the interaction terms will be handled in the fermionic-field path integral formalism \cite{cite_50}. For this, we pass into the Grassmann representation for the fermionic variables, and we write the partition function of the system in the imaginary time fermion path integral method. We introduce the imaginary-time variables $\tau$, at each lattice site ${\bf{r}}$ and the variables $\tau$ vary in the interval $(0,\beta)$, where $\beta=1/T$ with T being the temperature. The grand canonical partition function of the system is
\begin{eqnarray}
Z=\int\left[D\bar{X}DX\right]\left[D\bar{Y}DY\right]e^{-S\left[\bar{X},X,\bar{Y},Y\right]},
\label{Equation_5}
\end{eqnarray}
and the fermionic action $S\left[\bar{X},X,\bar{Y},Y\right]$ is given as follows
\begin{eqnarray}
S\left[\bar{X},X,\bar{Y},Y\right]=\sum_{l=1,2}S^{(l)}_{\rm B}\left[\bar{X},X\right]
\nonumber\\
+\sum_{l=1,2}S^{(l)}_{\rm B}\left[\bar{Y},Y\right]+\int^{\beta}_{0}d\tau H\left(\tau\right).
\label{Equation_6}
\end{eqnarray}
Here, the first two terms are the Berry terms for the layers with the indices $\ell=1,2$
\begin{eqnarray}
S^{(l)}_{\rm B}\left[\bar{X},X\right]=\sum_{{\bf{r}},\sigma}\int^{\beta}_{0}d\tau \bar{X}_{l,\sigma}({\bf{r}},\tau)\frac{\partial}{\partial \tau}X_{l,\sigma}({\bf{r}},\tau),
\label{Equation_7}
\newline\\
S^{(l)}_{\rm B}\left[\bar{Y},Y\right]=\sum_{{\bf{r}},\sigma}\int^{\beta}_{0}d\tau \bar{Y}_{l,\sigma}({\bf{r}},\tau)\frac{\partial}{\partial \tau}Y_{l,\sigma}({\bf{r}},\tau),
\label{Equation_8}
\end{eqnarray}
where we have introduced the following notation for the fermionic operators: $X_{1,\sigma}({\bf{r}},\tau)=a_{1,\sigma}({\bf{r}},\tau)$, $X_{2,\sigma}({\bf{r}},\tau)=\tilde{a}_{2,\sigma}({\bf{r}},\tau)$, $Y_{1,\sigma}({\bf{r}},\tau)=b_{1,\sigma}({\bf{r}},\tau)$ and $Y_{2,\sigma}({\bf{r}},\tau)=\tilde{b}_{2,\sigma}({\bf{r}},\tau)$. The Hamiltonian $H\left(\tau\right)$ of the BLG system, in the last term in Eq.(\ref{Equation_6}), is described in Eq.(\ref{Equation_1}), and here we will write it in the more convenient form
\begin{eqnarray}
H=-\gamma_0\sum_{\substack{\left\langle {\bf{r}},{\bf{r}}'\right\rangle,\\ \sigma}}\left(a_{1,\sigma}({\bf{r}}\tau)b_{1{\bf{r}}',\sigma}({\bf{r}}'\tau)+h.c.\right)
\nonumber\\
-\gamma_0\sum_{\substack{\left\langle {\bf{r}},{\bf{r}}'\right\rangle,\\ \sigma}}\left(\bar{\tilde{a}}_{2,\sigma}({\bf{r}},\tau)\tilde{b}_{2,\sigma}({\bf{r}},\tau)+h.c.\right)
\nonumber\\
-\gamma_1\sum_{{\bf{r}},\sigma}\left(\bar{{b}}_{1,\sigma}({\bf{r}},\tau)\tilde{a}_{2,\sigma}({\bf{r}},\tau)+h.c.\right)
\nonumber\\
+U\sum_{\substack{l,\\ \eta=X,Y}}\left[\frac{\left({n^{\eta}_{l}}({\bf{r}},\tau)\right)^{2}}{4}-\left({S^{\eta}_{l,z}}({\bf{r}},\tau)\right)^{2}\right]
\nonumber\\
-\mu_{1}\sum_{{\bf{r}},\sigma}n^{a}_{1,\sigma}({\bf{r}},\tau)-\mu_{2}\sum_{{\bf{r}},\sigma}n^{b}_{1,\sigma}({\bf{r}}\tau)-\mu_{2}\sum_{{\bf{r}},\sigma}n^{\tilde{a}}_{2,\sigma}({\bf{r}},\tau)
\nonumber\\
-\mu_{1}\sum_{{\bf{r}},\sigma}n^{\tilde{b}}_{2,\sigma}({\bf{r}},\tau)-V\sum_{{\bf{r}},\sigma,\sigma'}|\chi_{{\bf{r}},\sigma\sigma'}(\tau)|^{2}.
\label{Equation_9}
\end{eqnarray}
We have introduced in Eq.(\ref{Equation_9}) the $z$-component of the generalized spin operator ${\bf{S}}^{\eta}_{l}({\bf{r}},\tau)=1/2\sum_{\alpha,\beta = \uparrow, \downarrow}\bar{\eta}_{l,\alpha}({\bf{r}}\tau)\hat{\sigma}_{\alpha\beta}\eta_{l,\beta}({\bf{r}},\tau)$, for different sublattices in the layers of the BLG structure. It is defined as $S^{\eta}_{l,z}({\bf{r}},\tau)=1/2\left(\eta_{l,\uparrow}({\bf{r}},\tau)-\eta_{l,\downarrow}({\bf{r}},\tau)\right)$. The chemical potentials $\mu_{1}$ and $\mu_{2}$ in Eq.(\ref{Equation_9}) are the shifted chemical potentials, defined as: $\mu_{1}=\mu+U/2$, $\mu_{2}=\mu+U/2+V$. Indeed, the chemical potentials of electrons on the nonequivalent sublattice sites get different shifts in different layers due to the stacking ordering of the BLG structure. We have introduced the new complex variables $\chi_{\sigma\sigma'}({\bf{r}},\tau)$ and their complex conjugates $\bar{\chi}_{\sigma\sigma'}({\bf{r}},\tau)$ in the last term in Eq.\ref{Equation_9}), where $\chi_{\sigma\sigma'}({\bf{r}},\tau)$ is defined as 
\begin{eqnarray}
\chi_{\sigma\sigma'}({\bf{r}},\tau)=\bar{b}_{1,\sigma}({\bf{r}},\tau)\tilde{a}_{2,\sigma}({\bf{r}},\tau).
\label{Equation_10}
\end{eqnarray}
The Hamiltonian, in the form given in Eq.(\ref{Equation_9}), is more suitable for further decouplings of four-fermionic terms, within the Hubbard-Stratanovich saddle-point linearisation procedure.
We give here the procedure of the real-space linearization of the $U-V$ terms for the case of the $a$-sublattice in the layer-1 in the BLG. Namely, for the $U$-term in Eq.(\ref{Equation_9}) we have
\begin{eqnarray}
&&e^{-U/4\sum_{{\bf{r}}}\int^{\beta}_{0}d\tau\left(n^{a}_{1}({\bf{r}},\tau)-\frac{2\mu_{1}}{U}\right)^{2}}\sim
\nonumber\\
&&\sim \int\left[DV^{a}_{1}\right]e^{\sum_{i}\int^{\beta}_{0}d\tau\left[-\left(\frac{V^{a}_{1}({\bf{r}},\tau)}{\sqrt{U}}\right)^{2}+iV^{a}_{1}({\bf{r}},\tau)\left(n^{a}_{1}({\bf{r}},\tau)-\frac{2\mu_{1}}{U} \right)\right]}.
\nonumber\\
\label{Equation_11}
\end{eqnarray}
The integral in the right hand side (r.h.s.), in Eq.(\ref{Equation_11}), is over the decoupling field variables $V^{a}_{1}({\bf{r}},\tau)$ (which are introduced at each sublattice site position ${\bf{r}}$ and at each time $\tau$) coupled to the density term $n^{a}_{1}({\bf{r}},\tau)$. The field integral, in r.h.s., in Eq.(\ref{Equation_11}), can be evaluated by the steepest descent method. We get
\begin{eqnarray}
\int\left[DV^{a}_{1}\right]e^{\sum_{{\bf{r}}}\int^{\beta}_{0}d\tau\left[-\left(\frac{V^{a}_{1}({\bf{r}},\tau)}{\sqrt{U}}\right)^{2}+iV^{a}_{1}({\bf{r}},\tau)\left(n^{a}_{1}({\bf{r}},\tau)-\frac{2\mu_{1}}{U} \right)\right]}\sim
\nonumber\\
\sim e^{-U/2\sum_{{\bf{r}}}\int^{\beta}_{0}d\tau \left(\bar{n}^{a}_{1}-\frac{2\mu_{1}}{U}\right)\left(n^{a}_{1}({\bf{r}},\tau)-\frac{2\mu_{1}}{U} \right)}.
\nonumber\\
\label{Equation_12}
\end{eqnarray}
Here, in order to obtain the r.h.s. in Eq.(\ref{Equation_12}), we have replaced the field integration over $V^{a}_{1}({\bf{r}},\tau)$ by the value of the function in the exponential at the saddle-point value of the decoupling potential $\upsilon^{a}_{1}=iU/2\left(\bar{n}^{a}_{1}-2\mu_{1}/U\right)$, where the average density $\bar{n}^{a}_{1}$ is defined with the help of the total action of the system, given in Eq.(\ref{Equation_6}). Thus, we have $\bar{n}^{a}_{1}=\left\langle n^{a}_{1,\uparrow}({\bf{r}},\tau)+n^{a}_{1,\downarrow}({\bf{r}},\tau)\right\rangle$. 

The same procedure could be repeated also for the nonlinear density terms on the other sublattices in the BLG structure, containing $n^{\tilde{a}}_{2}({\bf{r}})$, $n^{b}_{1}({\bf{r}})$ and $n^{\tilde{b}}_{2}({\bf{r}})$ density terms. The decoupling of the nonlinear density-difference term, in Eq.(\ref{Equation_9}), is also straightforward. Namely, for the $l$-layer and $\eta$-type sublattice variables we have
  \begin{widetext}
\begin{eqnarray}
e^{U\sum_{{\bf{r}}}\int^{\beta}_{0}d\tau \left(S^{\eta}_{l,z}({\bf{r}},\tau)\right)^{2}}=e^{U/4\sum_{{\bf{r}}}\int^{\beta}_{0}d\tau \left(n^{\eta}_{l,\uparrow}({\bf{r}},\tau)-n^{\eta}_{l,\downarrow}({\bf{r}},\tau)\right)^{2}}\sim
\nonumber\\
\sim \int{\left[D\Delta^{\eta}_{c,l}\right]}e^{\sum_{{\bf{r}}}\int^{\beta}_{0}d\tau \left[-\left(\frac{\Delta^{\eta}_{c,l}({\bf{r}},\tau)}{\sqrt{U}}\right)^{2}+\Delta^{\eta}_{c,l}({\bf{r}},\tau)\left(n^{\eta}_{l,\uparrow}({\bf{r}},\tau)-n^{\eta}_{l,\downarrow}({\bf{r}},\tau)\right)\right]}.
\label{Equation_13}
\end{eqnarray}
  \end{widetext}
The saddle-point values of the variables $\Delta^{\eta}_{c,l}({\bf{r}},\tau)$ are given by $\delta^{\eta}_{c,l}=U/2\left\langle n^{\eta}_{l,\uparrow}({\bf{r}},\tau)-n^{\eta}_{l,\downarrow}({\bf{r}},\tau)\right\rangle$. Thus, it is proportional to the difference between the electron densities with the opposite spin polarizations. For simplicity, we suppose the case of the spin balanced BLG layers, with equal density numbers for each spin direction, i.e. $\left\langle n^{\eta}_{l,\uparrow}({\bf{r}},\tau)\right\rangle=\left\langle n^{\eta}_{l,\downarrow}({\bf{r}},\tau)\right\rangle$ and the quantities $\delta^{\eta}_{c,l}$ vanish in the case. For the case of the half-filling occupation, considered here, we put $\left\langle n^{\eta}_{l,\sigma}({\bf{r}},\tau)\right\rangle=1/2$, for each spin direction $\sigma=\uparrow, \downarrow$. 

Next, for decoupling the last term interaction $V$-term in Eq.(\ref{Equation_9}), we apply the complex form of the Hubbard-Stratanovich transformation \cite{cite_50} for the one-component fermion-field
  \begin{eqnarray}
  &&e^{V\sum_{{\bf{r}},\sigma,\sigma'}\int^{\beta}_{0}d\tau|\chi_{\sigma\sigma'}({\bf{r}},\tau)|^{2}}=
  \nonumber\\
 &&=\int{\left[D\bar{\Gamma}D\Gamma\right]}e^{\sum_{{\bf{r}}}\int^{\beta}_{0}d\tau -\frac{|\Gamma_{\sigma\sigma'}({\bf{r}},\tau)|^{2}}{V}}\times
\nonumber\\ 
&&\times e^{\sum_{{\bf{r}}}\int^{\beta}_{0}d\tau \bar{\Gamma}_{\sigma\sigma'}({\bf{r}},\tau)\chi_{\sigma\sigma'}({\bf{r}},\tau)+\bar{\chi}_{\sigma\sigma'}({\bf{r}},\tau){\Gamma}_{\sigma\sigma'}({\bf{r}},\tau)}.
\label{Equation_14}
  \end{eqnarray}
In fact, the saddle-point value of the decoupling field $\Gamma_{\sigma\sigma'}({\bf{r}},\tau)$, introduced in Eq.(\ref{Equation_14}), is directly related to the excitonic gap parameter. Indeed, we have 
  \begin{eqnarray}
  \Delta_{\sigma\sigma'}=V\left\langle \bar{b}_{1,\sigma}({\bf{r}},\tau)\tilde{a}_{2,\sigma'}({\bf{r}},\tau)\right\rangle.
  \label{Equation_15}
   \end{eqnarray}
We consider here the homogeneous BLG structure with the pairing between the particles with the same orientation of spin variables, i.e. $\Delta_{\sigma\sigma'}=\Delta_{\sigma}\delta_{\sigma\sigma'}$.
Furthermore, we can write the total action of the system in the Fourier representation, given by the transformations $\eta_{l,\sigma}({\bf{r}},\tau)=\frac{1}{\beta{N}}\sum_{{\bf{k}},\nu_{n}}\eta_{{\bf{k}},\sigma}(\nu_{n})e^{i\left({\bf{k}}{\bf{r}}_{i}-\nu_{n}\tau\right)}$, where $\nu_{n}=\pi\left(2n+1\right)/\beta$, with $n=0,\pm1,\pm2,...$, are the fermionic Matsubara frequencies \cite{cite_51}, and $N$ is the total number of sites on the $\eta$-type sublattice, in the layer $l$. We introduce the four component Nambu-spinors at each discrete state ${\bf{k}}$ in the reciprocal space and for each spin direction $\sigma=\uparrow, \downarrow$, ${\psi}_{{\bf{k}},\sigma}(\nu_{n})=\left[a_{1{\bf{k}},\sigma},b_{1{\bf{k}},\sigma},\tilde{a}_{2{\bf{k}},\sigma},\tilde{b}_{2{\bf{k}},\sigma}\right]^{T}$. Then the action of the system reads as
 \begin{eqnarray} S\left[\bar{\psi},\psi,\bar{\Delta},\Delta\right]=\frac{1}{\beta{N}}\sum_{{\bf{k}},\sigma}\bar{\psi}_{{\bf{k}},\sigma}(\nu_{n})G^{-1}_{{\bf{k}},\sigma}(\nu_{n}){\psi}_{{\bf{k}},\sigma}(\nu_{n}).
  \nonumber\\
  \label{Equation_16}
   \end{eqnarray}
Here, $G^{-1}_{{\bf{k}},\sigma}(\nu_{n})$, is the inverse Green's function matrix, of size $4\times4$. It is defined as
\begin{eqnarray}
\footnotesize
\arraycolsep=0pt
\medmuskip = 0mu
{G}^{-1}_{{\bf{k}},\sigma}\left(\nu_{n}\right)=\left(
\begin{array}{ccccrrrr}
E_{1}(\nu_{n}) & -\tilde{\gamma}_{1{\bf{k}}} & 0 & 0\\
-\tilde{\gamma}^{\ast}_{1{\bf{k}}} &E_{2}(\nu_{n})  & -\gamma_{1}-\bar{\Delta}_{\sigma} & 0 \\
0 & -\gamma_{1}-{\Delta}_{\sigma} & E_{2}(\nu_{n}) & -\tilde{\gamma}_{2{\bf{k}}} \\
0 & 0 & -\tilde{\gamma}^{\ast}_{2{\bf{k}}} & E_{1}(\nu_{n}) 
\end{array}
\right).
\label{Equation_17}
\end{eqnarray}

The diagonal elements of the matrix in Eq.(\ref{Equation_17}) are the energy parameters $E_{1}(\nu_{n})=-i\nu_{n}-\mu^{\rm eff}_{1}$ and $E_{2}(\nu_{n})=-i\nu_{n}-\mu^{\rm eff}_{2}$, where, the effective chemical potentials $\mu^{\rm eff}_{1}$ and $\mu^{\rm eff}_{2}$, are defined with the help of the intralayer and interlayer interaction parameters $U$ and $V$ as 
\begin{eqnarray}
\mu^{\rm eff}_{1}=\mu+U/4,
\label{Equation_18}
\newline\\
\mu^{\rm eff}_{2}=\mu+U/4+V.
\label{Equation_19}
\end{eqnarray}
Thus the effect of the interaction parameters is such that the chemical potential $\mu$ gets shifted in BLG system and also the position of the CNP point gets shifted, as it was discussed in Ref.\onlinecite{cite_11}.
The parameters $\tilde{\gamma}_{l{\bf{k}}}$, in Eq.(\ref{Equation_17}), $l=1,2$, are the renormalized (nearest neighbors) hopping amplitudes $\tilde{\gamma}_{l{\bf{k}}}=z\gamma_{l{\bf{k}}}t$, where the ${\bf{k}}$-dependent parameters $\gamma_{1{\bf{k}}}$ and $\gamma_{2{\bf{k}}}$ are the energy dispersions in the BLG layers with $l=1$ and $l=2$, respectively. We have $\gamma_{1{\bf{k}}}=1/z\sum_{\bm{\mathit{\delta}}}e^{-i{{\bf{k}}\bm{\mathit{\delta}}}}$ for $\ell=1$ (and $\gamma_{2{\bf{k}}}=1/z\sum_{\bm{\mathit{\delta}}'}e^{-i{{\bf{k}}\bm{\mathit{\delta}}'}}$ for the layer with $l=2$). The parameter $z$ is the number of the nearest neighbors lattice sites on the honeycomb lattice for a given sublattice variable and $z=3$ for each monolayer (see in Fig.~\ref{fig:Fig_1}). The vectors $\bm{\mathit{\delta}}$ and $\bm{\mathit{\delta}}'$ are the nearest neighbor vectors in different layers. The components of $\bm{\mathit{\delta}}$, for the bottom layer-1, are given by $\bm{\mathit{\delta}}_{1}=\left({a_{0}}/{2\sqrt{3}},a_{0}/2\right)$, $\bm{\mathit{\delta}}_{2}=\left({a_{0}}/{2\sqrt{3}},-a_{0}/2\right)$, $\bm{\mathit{\delta}}_{3}=\left(-a_{0}/\sqrt{3},0\right)$, and $a_{0}=\sqrt{3}a$ is the sublattice constant (with $a$, being the carbon-carbon length in the graphene sheets). In the layer-2, we have$bm{\mathit{\delta}}'_{1}=\left(a_{0}/\sqrt{3},0\right)$, $\bm{\mathit{\delta}}'_{2}=\left(-{a_{0}}/{2\sqrt{3}},-a_{0}/2\right)$, $\bm{\mathit{\delta}}'_{3}=\left(-{a_{0}}/{2\sqrt{3}},a_{0}/2\right)$. It is not difficult to realise that $\bm{\mathit{\delta}}'=-\bm{\mathit{\delta}}$. Then, for the function $\gamma_{1{\bf{k}}}$, we have $\gamma_{1{\bf{k}}}=1/3\left(e^{-ik_{x}a}+2e^{i\frac{k_{x}a}{2}}\cos{\frac{\sqrt{3}}{2}k_{y}a}\right)$, where $a$ is the carbon-carbon interatomic distance. By the convention, we put $a\equiv 1$, for both layers. For a given geometry of the AB-stacked BLG it is not difficult to realize that $\gamma_{2{\bf{k}}}=\gamma^{\ast}_{1{\bf{k}}}\equiv\gamma^{\ast}_{{\bf{k}}}$ and it follows that $\tilde{\gamma}_{2{\bf{k}}}=\tilde{\gamma}^{\ast}_{1{\bf{k}}}\equiv\tilde{\gamma}^{\ast}_{{\bf{k}}}$, where we have omitted the layer index $l$.   

We assume here that the pairing gap is real and is not spin-dependent ($\Delta_\sigma\equiv\Delta=\bar{\Delta}$). Therefore, the structure of the Green's function matrix does not changes for the opposite spin direction: $\hat{G}^{-1}_{{\bf{k}},\uparrow}\left(\nu_{n}\right)\equiv \hat{G}^{-1}_{{\bf{k}},\downarrow}\left(\nu_{n}\right)$. Next, we use the half-filling condition in each layer of the BLG system which determine the chemical potential in the BLG. For the layer-$\ell$, this condition holds that $\bar{n}^{a}_{\ell}+\bar{n}^{b}_{\ell}=1$. Here, we present the resulting system of coupled nonlinear self-consistent equations for the chemical potential $\mu$ and excitonic pairing gap parameter $\Delta$. We get 
\begin{eqnarray}
\frac{4}{N}\sum_{{\bf{k}}}\sum_{i=1,..,4}\alpha_{i{{\bf{k}}}}n_{\rm F}(\mu-\varepsilon_{i{\bf{k}}})=1,
\label{Equation_20}
\newline\\
\Delta=\frac{V(\gamma_{1}+\Delta)}{N}\sum_{{\bf{k}}}\sum_{i=1,..,4}\beta_{i{{\bf{k}}}}n_{\rm F}(\mu-\varepsilon_{i{\bf{k}}}),
\label{Equation_21}
\end{eqnarray}
where the dimensionless coefficients $\alpha_{i{{\bf{k}}}}$, in Eq.(\ref{Equation_20}) with $i=1,..4$, are given as
\begin{eqnarray}
\footnotesize
\arraycolsep=0pt
\medmuskip = 0mu
\alpha_{i{{\bf{k}}}}=(-1)^{i+1}
\left\{
\begin{array}{cc}
\displaystyle  & \frac{\varepsilon^{3}_{i{\bf{k}}}+a_{1{\bf{k}}}\varepsilon^{2}_{i{\bf{k}}}+a_{2{\bf{k}}}\varepsilon_{i{\bf{k}}}+a_{3\bf{k}}}{\left(\varepsilon_{1{\bf{k}}}-\varepsilon_{2{\bf{k}}}\right)\left(\varepsilon_{i{\bf{k}}}-\varepsilon_{3{\bf{k}}}\right)\left(\varepsilon_{i{\bf{k}}}-\varepsilon_{4{\bf{k}}}\right)},  \ \ \  $if$ \ \ \ i=1,2,
\newline\\
\newline\\
\displaystyle  & \frac{\varepsilon^{3}_{i{\bf{k}}}+a_{1{\bf{k}}}\varepsilon^{2}_{i{\bf{k}}}+a_{2{\bf{k}}}\varepsilon_{i{\bf{k}}}+a_{3{\bf{k}}}}{\left(\varepsilon_{3{\bf{k}}}-\varepsilon_{4{\bf{k}}}\right)\left(\varepsilon_{i{\bf{k}}}-\varepsilon_{1{\bf{k}}}\right)\left(\varepsilon_{i{\bf{k}}}-\varepsilon_{2{\bf{k}}}\right)},  \ \ \  $if$ \ \ \ i=3,4
\end{array}\right.
\nonumber\\
\label{Equation_22}
\end{eqnarray}
with 
\begin{eqnarray}
a_{1{\bf{k}}}=-2\mu^{\rm eff}_{2}-\mu^{\rm eff}_{1},
\label{Equation_23} 
\newline\\
a_{2{\bf{k}}}=\mu^{\rm eff}_{1}\left(\mu^{\rm eff}_{2}+2\mu^{\rm eff}_{1}\right)-\Delta^{2}-|\tilde{\gamma}_{{\bf{k}}}|^{2}, 
\label{Equation_24}
\end{eqnarray}
and
\begin{eqnarray}
a_{3{\bf{k}}}=-\mu^{\rm eff}_{1}\left(\mu^{\rm eff}_{2}\right)^{2}+\mu^{\rm eff}_{1}\Delta^{2}+\mu^{\rm eff}_{2}|\tilde{\gamma}_{{\bf{k}}}|^{2}.
\label{Equation_25}
\end{eqnarray}
The coefficients $\beta_{i{{\bf{k}}}}$ in Eq.(\ref{Equation_21}), with $i=1,..4$ are given by the relations
 \begin{eqnarray}
\beta_{i{{\bf{k}}}}=
\left\{
\begin{array}{cc}
\displaystyle  & \frac{(-1)^{i+1}\left(\mu^{\rm eff}_{1}-\varepsilon_{i{\bf{k}}}\right)^{2}}{\left(\varepsilon_{1{\bf{k}}}-\varepsilon_{2{\bf{k}}}\right)\left(\varepsilon_{i{\bf{k}}}-\varepsilon_{3{\bf{k}}}\right)\left(\varepsilon_{i{\bf{k}}}-\varepsilon_{4{\bf{k}}}\right)},  \ \ \  $if$ \ \ \ i=1,2,
\newline\\
\newline\\
\displaystyle  & \frac{(-1)^{i}\left(\mu^{\rm eff}_{1}-\varepsilon_{i{\bf{k}}}\right)^{2}}{\left(\varepsilon_{3{\bf{k}}}-\varepsilon_{4{\bf{k}}}\right)\left(\varepsilon_{i{\bf{k}}}-\varepsilon_{1{\bf{k}}}\right)\left(\varepsilon_{i{\bf{k}}}-\varepsilon_{2{\bf{k}}}\right)},  \ \ \  $if$ \ \ \ i=3,4 .
\end{array}\right.
\nonumber\\
\label{Equation_26}
\end{eqnarray}
The function $n_{F}\left(x\right)$, in Eqs.(\ref{Equation_20}) and (\ref{Equation_21}), is the Fermi-Dirac distribution function $n_{F}\left(x\right)=1/\left(e^{\beta(x-\mu)}+1\right)$.  
The energy parameters $\varepsilon_{i{\bf{k}}}$ define the interacting band structure in the BLG in our problem of the excitonic effects in the BLG. They are given by the following relations
\begin{widetext}
\begin{eqnarray}
\varepsilon_{1,2{\bf{k}}}=-\frac{1}{2}\left[\Delta+\gamma_{1}\pm\sqrt{\left(V-\Delta-\gamma_{1}\right)^{2}+4|\tilde{\gamma}_{{\bf{k}}}|^{2}}\right]+\bar{\mu},
\label{Equation_27}
\newline\\
\varepsilon_{3,4{\bf{k}}}=-\frac{1}{2}\left[-\Delta-\gamma_{1}\pm\sqrt{\left(V+\Delta+\gamma_{1}\right)^{2}+4|\tilde{\gamma}_{{\bf{k}}}|^{2}}\right]+\bar{\mu}.
\label{Equation_28}
\end{eqnarray}
\newline\\
\end{widetext}
The exact numerical solution of Eqs.(\ref{Equation_20})-(\ref{Equation_21}), and the changes in the electronic band structure of BLG system, in the presence of the excitonic pairing, are discussed in Refs.\onlinecite{cite_11, cite_52}. The parameter $\bar{\mu}$ plays the crucial role in the whole physics related to the excitonic BLG. It plays the role of the exact Fermi energy in the BLG and it is defined with the help of the effective chemical potentials as $\bar{\mu}=1/2\left(\mu^{\rm eff}_{1}+\mu^{\rm eff}_{2}\right)$. As the results show, the Fermi energy in the bilayer graphene as a function of the interaction parameter $V$ and at $T = 0$ has also a very large jump at the CNP, similar to the exact chemical potential in BLG, and it happens nearly at the same value of the interlayer interaction parameter $V=1.49\gamma_0$. Moreover, at the zero interlayer interaction limit, the bare chemical potential $\bar{\mu}$ coincides with the Dirac's crossing energy level $\varepsilon_{\cal{D}}$ \cite{cite_11}. The Coulomb interaction parameter $U$ redefines the Fermi level in the BLG by the way that $\varepsilon_{F}=\bar{\mu}=\mu+\kappa{U}+0.5V$ (with $\kappa=0.25$) \cite{cite_11, cite_12} and the chemical potential $\mu$ could be calculated self-consistently (see in Ref. \onlinecite{cite_11}).  
%
\subsection{\label{sec:Section_2_2} The effect of the electric field}
%
We suppose that the bilayer graphene gets excited in the external electromagnetic field, with the electric field component polarized along the $x$-axis in the graphene's sheet. Then, in order to consider the electric current response of the system, we include the vector potential ${\bf{A}}({\bf{r}})$ in the tight binding part of the Hamiltonian in Eq.(\ref{Equation_2}). This could be done via the Peierls-Onsager substitution \cite{cite_48, cite_49}
\begin{eqnarray}
 C^{\dag}_{\ell\sigma}({\bf{r}})C_{\ell\sigma}({\bf{r}}')\rightarrow C^{\dag}_{\ell\sigma}({\bf{r}})e^{-\frac{ie}{\hbar{c}}\int^{{\bf{r}}}_{{\bf{r}}'}{\bf{A}}({\bf{l}})d{\bf{l}}}C_{\ell\sigma}({\bf{r}}')
 	\label{Equation_29}
 	\end{eqnarray}
with $C_{1}=a,b$ for $\ell=1$, and $C_{2}=\tilde{a},\tilde{b}$ for $\ell=2$. The electron operators in the interaction terms do not get modified in that case because of the local nature of those terms. The shift of the operators product in Eq.(\ref{Equation_29}) by a phase factor is related to the electronic Wannier states modifications in that case (for a detailed description, see in Ref.\onlinecite{cite_49}). After inserting the transformations, given in Eq.(\ref{Equation_29}), into the Hamiltonian in Eq.(\ref{Equation_1}) and then expanding it up to first order in vector potential ${\bf{A}}({\bf{r}})$, we get 
    \begin{eqnarray}
    H'=H-\frac{1}{c}\sum_{\bf{r},i}\sum_{\ell}{{A}_{\ell{i}}}({\bf{r}}){j}_{{\ell}i}({\bf{r}}),
        \label{Equation_30}
    \end{eqnarray}
where $c$ is the speed of light and ${j}_{{\ell}i}({\bf{r}})$ is the component of the current operator along the direction $i$, in the given sheet $\ell$. It could be obtained from the expression in Eq.(\ref{Equation_30}), after the functional differentiation of the Hamiltonian with respect to $A_{{\ell}i}(\mathbf{r})$ (here $i$ indicates the component of the vector potential ${\bf{A}}$, along the considered direction $i$). For the total current density operator in the BLG, we get straightforwardly
           \begin{eqnarray}
          &&j_{i}({\mathbf{r}})=-\sum_{\ell}\frac{\delta{H'}}{\delta{\left(A_{{\ell}i}(\mathbf{r})/c\right)}}=
					\nonumber\\
					&&-\frac{ie\gamma_0}{\hbar}\sum_{\bm{\mathit{\delta}}}\left[a^{\dag}_{\sigma}(\mathbf{r}+\delta)b_{\sigma}(\mathbf{r})\bm{\mathit{\delta}}_{i}-h.c.\right]
          \nonumber\\
          &&+\frac{ie\gamma_0}{\hbar}\sum_{\bm{\mathit{\delta}}'}\left[\tilde{a}^{\dag}_{\sigma}(\mathbf{r}+\delta)\tilde{b}_{\sigma}(\mathbf{r})\bm{\mathit{\delta}}'_{i}-h.c.\right].
           \label{Equation_31}
           \end{eqnarray}
We consider here only the linear contribution of the external vector potential into the Hamiltonian in Eq.(\ref{Equation_1}). This term is responsible for the paramagnetic part of the current operator, and we neglected the second order diamagnetic contribution in Eq.(\ref{Equation_31}).
%
\section{\label{sec:Section_3} The UV optical conductivity}
%
 \subsection{\label{sec:Section_3_1} The polarization function}
 %
 We will consider the retarded polarization function (see in the Appendix \ref{sec:Section_7}) in order to calculate the optical conductivity in the system. In the continuum approximation at the $K$ point, we have for the velocity operator $|v_{\mathbf{k}x}|^{2}=v^{2}_{F}$ (see also in Ref.\onlinecite{cite_20}), where $v_{F}$ is the Fermi velocity, which relates to the intralayer hopping parameter $\gamma_0$, i.e., $v_{F}=\sqrt{3}a_{0}\gamma_0/2\hbar$. For the $A$-sublattice in the bottom layer-1, we have defined the normal Green's functions as follows
 \begin{eqnarray}
 {\cal{G}}_{ aa}\left({\bf{k}}\tau,{\bf{k}}\tau'\right)=\frac{1}{\beta{N}}\left\langle a_{{\bf{k}}}(\tau)\bar{a}_{{\bf{k}}}(\tau')\right\rangle.
 \end{eqnarray}
 The similar expression could be written also for the $B$-sublattice Green's function $	{\cal{G}}_{ bb}\left({\bf{k}}\tau,{\bf{k}}\tau'\right)$. For the top layer sublattice Green's functions ${\cal{G}}_{ \tilde{a}\tilde{a}}$ and ${\cal{G}}_{ \tilde{b}\tilde{b}}$ we have the relations ${\cal{G}}_{ \tilde{a}\tilde{a}}={ \cal{G}}_{ bb}$ and ${\cal{G}}_{ \tilde{b}\tilde{b}}={\cal{G}}_{ aa}$ \cite{cite_52}. Then, for the component of the polarization function, along the current direction, we get
    \begin{eqnarray}
    \Pi_{ xx}(i\omega_m)=\frac{2e^{2}v^{2}_{F}}{\beta}\sum_{\mathbf{k}\nu_n}\left[{\cal{G}}_{ aa}(\mathbf{k},\nu_n){\cal{G}}_{ bb}(\mathbf{k},\nu_n-\omega_m)\right.
    \nonumber\\
   \left. +{\cal{G}}_{ \tilde{a}\tilde{a}}(\mathbf{k},\nu_n){\cal{G}}_{ \tilde{b}\tilde{b}}(\mathbf{k},\nu_n-\omega_m)\right.
   \nonumber\\
     \left.+{\cal{G}}_{ aa}(\mathbf{k},\nu_n){\cal{G}}_{ bb}(\mathbf{k},\nu_n+\omega_m)
     \right.
     \nonumber\\
     \left.+{\cal{G}}_{ \tilde{a}\tilde{a}}(\mathbf{k},\nu_n){\cal{G}}_{ \tilde{b}\tilde{b}}(\mathbf{k},\nu_n+\omega_m)\right].
      \label{Equation_32}
    \end{eqnarray}
The explicit analytical forms of the Fourier transformed single-particle Green's functions in Eq.(\ref{Equation_32}) could be obtained after the functional derivation techniques \cite{cite_52}
\begin{eqnarray}
{\cal{G}}_{ aa}(\mathbf{k},\nu_n)=\sum^{4}_{i=1}\frac{\alpha_{i{\bf{k}}}}{i\nu_{n}+\varepsilon_{i{\bf{k}}}},
\label{Equation_33}
\end{eqnarray}
where the energy the parameters $\varepsilon_{i{\bf{k}}}$ are defined in Eqs.(\ref{Equation_27}) and (\ref{Equation_28}). Here, $\varepsilon_{1{\bf{k}}}$ and $\varepsilon_{4{\bf{k}}}$ are the split valence and conduction bands, and $\varepsilon_{2{\bf{k}}}$ and $\varepsilon_{3{\bf{k}}}$ are the low energy conduction and valence bands, according to the usual tight-binding definitions \cite{cite_12, cite_40}. The coefficients $\alpha_{i{\bf{k}}}$, figuring in the nominator in the sum, in Eq.(\ref{Equation_22}). Next, for the $B$-sublattice Green function ${\cal{G}}_{ bb}(\mathbf{k},\nu_n)$, we have \cite{cite_52}
		\begin{eqnarray}
   	{\cal{G}}_{ bb}(\mathbf{k},\nu_n)=\sum^{4}_{i=1}\frac{\gamma_{i{\bf{k}}}}{i\nu_{n}+\varepsilon_{i{\bf{k}}}}.
   	\label{Equation_34}
   	\end{eqnarray}
The coefficients $\gamma_{i{{\bf{k}}}}$, in the nominators in Eq.(\ref{Equation_34}) are defined as
\begin{eqnarray}
\footnotesize
\arraycolsep=0pt
\medmuskip = 0mu
\gamma_{i{{\bf{k}}}}=(-1)^{i+1}
\left\{
\begin{array}{cc}
\displaystyle  & \frac{\varepsilon^{3}_{i{\bf{k}}}+a'_{1{\bf{k}}}\varepsilon^{2}_{i{\bf{k}}}+a'_{2{\bf{k}}}\varepsilon_{i{\bf{k}}}+a'_{3\bf{k}}}{\left(\varepsilon_{1{\bf{k}}}-\varepsilon_{2{\bf{k}}}\right)\left(\varepsilon_{i{\bf{k}}}-\varepsilon_{3{\bf{k}}}\right)\left(\varepsilon_{i{\bf{k}}}-\varepsilon_{4{\bf{k}}}\right)},  \ \ \  $if$ \ \ \ i=1,2,
\newline\\
\newline\\
\displaystyle  & \frac{\varepsilon^{3}_{i{\bf{k}}}+a'_{1{\bf{k}}}\varepsilon^{2}_{i{\bf{k}}}+a'_{2{\bf{k}}}\varepsilon_{i{\bf{k}}}+a'_{3{\bf{k}}}}{\left(\varepsilon_{3{\bf{k}}}-\varepsilon_{4{\bf{k}}}\right)\left(\varepsilon_{i{\bf{k}}}-\varepsilon_{1{\bf{k}}}\right)\left(\varepsilon_{i{\bf{k}}}-\varepsilon_{2{\bf{k}}}\right)},  \ \ \  $if$ \ \ \ i=3,4
\end{array}\right.
\nonumber\\
\label{Equation_35}
\end{eqnarray}
with the coefficients $a'_{i{\bf{k}}}$, $i=1,...3$, given as 
\begin{align}
	&&a'_{1{\bf{k}}}=-2\mu^{\rm eff}_{1}-\mu^{\rm eff}_{2},
	\nonumber\\
	&&a'_{2{\bf{k}}}=\mu^{\rm eff}_{1}\left(\mu^{\rm eff}_{1}+2\mu^{\rm eff}_{2}\right)-|\tilde{\gamma}_{{\bf{k}}}|^{2},
	\nonumber\\
	&&a'_{3{\bf{k}}}=-\mu^{\rm eff}_{2}\left(\mu^{\rm eff}_{1}\right)^{2}+\mu^{\rm eff}_{1}|\tilde{\gamma}_{{\bf{k}}}|^{2}.
	\label{Equation_36}
\end{align}
We already showed, in Ref.\onlinecite{cite_52}, that the semiconducting (or insulating) state could be reached from the semimetallic limit in the BLG, thus leading to the enhancement of the Bardeen-Cooper-Schrieffer (BCS)-Bose-Einstein-Condensate (BEC) type crossover mechanism in the interacting BLG system. A detailed description of such a transition is given in Ref.\onlinecite{cite_52}. Next, for the product of the Fourier transformed single-particle Green's functions in Eq.(\ref{Equation_32}), we get
     \begin{eqnarray}
     &&{\cal{G}}_{ aa}(\mathbf{k},\nu_n){\cal{G}}_{ bb}(\mathbf{k},\nu_n-\omega_m)
     \nonumber\\
     &&=\sum_{i,j=1}^{4}\frac{\alpha_{i\mathbf{k}}\gamma_{j{\mathbf{k}}}}{\left(i\nu_n+\varepsilon_{i\mathbf{k}}\right)\left(i\left(\nu_n-\omega_m\right)+\varepsilon_{j\mathbf{k}}\right)}.
     \label{Equation_37}
     	\end{eqnarray}
Furthermore, we perform the summation over the fermionic Matsubara frequencies $\nu_{n}$, in all terms in Eq.(\ref{Equation_32}). Then, we obtain for the polarization operator 
 \begin{eqnarray}
  \Pi_{ xx}(i\omega_m)={2e^{2}v^{2}_{F}}\sum_{\mathbf{k}}\sum_{i,j=1}^{4}\sum_{\zeta=\pm1}\frac{\alpha_{i\mathbf{k}}\gamma_{i\mathbf{k}}+\gamma_{j\mathbf{k}}\alpha_{i\mathbf{k}}}{\varepsilon_{j\mathbf{k}}-\varepsilon_{i\mathbf{k}}+i\zeta\omega_m}
  \nonumber\\
  \times\left[n_{F}(\mu-\varepsilon_{i\mathbf{k}})-n_{F}(\mu-\varepsilon_{j\mathbf{k}})\right].
  \label{Equation_38}
\end{eqnarray}
The retarded polarization function enters into the expression of the real part of the longitudinal conductivity function $\Re{\sigma}_{\rm xx}$ and could be obtained after the standard analytical continuation techniques into the real frequency axis of the upper-half complex semi-plane. This procedure is given in Eq.(\ref{Equation_A_2}), in the Appendix \ref{sec:Section_7}. Then, for calculating the imaginary part of the polarization function, we use the real line version of the Sokhotskii-Plemelj identity, i.e., 
   \begin{eqnarray}
   &&\frac{1}{\varepsilon_{j\mathbf{k}}-\varepsilon_{i\mathbf{k}}+\zeta\Omega+i\zeta\eta^{+}}
   \nonumber\\
   &&={\cal{P}}\frac{1}{\varepsilon_{j\mathbf{k}}-\varepsilon_{i\mathbf{k}}+\zeta\Omega}-i\pi\zeta\delta(\varepsilon_{j\mathbf{k}}-\varepsilon_{i\mathbf{k}}+\zeta\Omega),
   \label{Equation_39}
   \end{eqnarray}
where ${\cal{P}}$ denotes the Cauchy principal value, and the parameter $\zeta$ takes two values $\pm1$. The function $\delta(x)$ in Eq.(\ref{Equation_39}) is the Dirac's delta function. The summation over the reciprocal lattice vectors $\mathbf{k}$, in Eq.(\ref{Equation_38}), could be replaced by the integration over the continuous variables via the introduction of the density of states (DOS) function $\rho(x)$ for the non-interacting bilayer graphene sheets, i.e., $\sum_{\mathbf{k}}\ldots=\int dx\rho(x)...$. The DOS in the non-interacting graphene layer, is defined as
   \begin{eqnarray}
   \rho(x)=\sum_{{\bf{k}}}\delta(x-\gamma_{{\bf{k}}}). 
   \label{Equation_40}
   \end{eqnarray}
   Beyond the Dirac's approximation it could be analytically expressed \cite{cite_3, cite_54} as
   \begin{eqnarray}
   \footnotesize
   \arraycolsep=0pt
   \medmuskip = 0mu\rho(x)=\frac{2|x|}{\pi^{2}|\gamma_{0}|^{2}}
   \left\{
   \begin{array}{cc}
   \displaystyle  & \frac{1}{\sqrt{\varphi\left(|{x}/{\gamma_0}|\right)}}{\mathbf{K}}\left[\frac{4|x/\gamma_0|}{\varphi\left(|{x}/{\gamma_0}|\right)}\right],  \ \ \  0<|x|<\gamma_0,
   \newline\\
   \displaystyle  & \frac{1}{\sqrt{4|{{x}/{\gamma_0}}|}}{\mathbf{K}}\left[\frac{\varphi\left(|x/\gamma_0|\right)}{4|{x}/{\gamma_0}|}\right],  \ \ \ \gamma_0<|x|<3\gamma_0,
   \end{array}\right.
   \nonumber\\
   \label{Equation_41}
   \end{eqnarray}
   where ${\mathbf{K}}(x)$ is the Elliptic integral of the first kind \cite{cite_55}  ${\mathbf{K}}(x)=\int^{\pi/2}_{0}dt/\sqrt{1-x^{2}\sin^{2}t}$ .
	The function $\varphi(x)$, in Eq.(\ref{Equation_41}), is given as \cite{cite_54} 
   \begin{eqnarray}
   \varphi(x)=\left(1+x\right)^{2}-\frac{\left(x^{2}-1\right)^{2}}{4}.
   \label{Equation_42}
   \end{eqnarray}
   For the real part of the optical conductivity function, we obtain finally
   \begin{eqnarray}
   &&\Re\sigma_{ xx}(\Omega)=\frac{\Im \Pi_{ xx}(\Omega)}{\Omega}=
   \nonumber\\
   &&{4\pi e^{2}v^{2}_{F}}\sum_{i,j=1}^{4}\int dx \rho(x)P_{ ij}(x)\delta\left[\Omega+\varepsilon_{j}(x)-\varepsilon_{i}(x)\right]\times
    \nonumber\\
   &&\times\left[n_{F}(\mu+\Omega-\varepsilon_{i}(x))-n_{F}(\mu-\varepsilon_{i}(x))\right].
    \label{Equation_43}
\end{eqnarray}
 The function $P_{ ij}(x)$, in Eq.(\ref{Equation_43}), is the index-permutation function, given as    
\begin{eqnarray}
P_{ij}(x)=\alpha_{i}(x)\gamma_{j}(x)+\alpha_{j}(x)\gamma_{i}(x),
\label{Equation_44}
\end{eqnarray} 
where the coefficients $\alpha_{i}(x)$ and $\gamma_{i}(x)$ are the continuous versions of the coefficients, given in Eqs.(\ref{Equation_22}) and (\ref{Equation_35}), above. 
%
 \subsection{\label{sec:Section_3_2} The optical conductivity}
 %
  In order to perform explicitly the integration in Eq.(\ref{Equation_43}), we will use the following rule for the composite Dirac's function: $\delta\left[f(x)\right]=\sum_{n}\delta(x-x_n)/|f'(x_n)|$, where $f(x)$ is a continuously differentiable function and $x_n$ are the solutions of the algebraic equation $f(x)=0$. Our calculations show obviously that the intraband optical transitions $(1\rightleftharpoons 3)$ and $(2\rightleftharpoons 4)$ do not contribute to the total optical conductivity function $\Re\sigma_{ xx}$, i.e., $\Re\sigma^{(13)}_{ xx}=\Re\sigma^{(31)}_{ xx}=\Re\sigma^{(24)}_{ xx}=\Re\sigma^{(42)}_{ xx}\equiv 0$. This effect leads furthermore (see in the Section \ref{sec:Section_4}) to the total suppression of the Drude region in the spectrum of the optical conductivity function. 
  
Then, after some calculations, we get the analytical expression for the real part of the conductivity function, in which only the interband transitions contribute (here, we normalize conductivity function in units of $\sigma_{ Bi}=e^{2}/2\hbar$, i.e., which is twice of the dc conductivity in the monolayer graphene $\sigma_{ MG}=e^{2}/4\hbar$ \cite{cite_20, cite_56}). After performing the explicit analytical integration in Eq.(\ref{Equation_43}), we get 
  \begin{widetext}
  	\begin{eqnarray}
  	&&\frac{\Re\sigma_{ xx}(\Omega)}{\sigma_{ Bi}}=\frac{6\gamma^{2}_{0}a^{2}}{\hbar\Omega}\left[ \Theta\left(-\Omega\right)\Theta\left(\Omega^{2}-b^{2}_{1}\right)\frac{{\cal{F}}_{12}\left(\Omega\right)}{|\Lambda_{12}(\xi^{ 12}_{1})|}+\Theta\left(\Omega\right)\Theta\left(\Omega^{2}-b^{2}_{1}\right)\frac{{\cal{F}}_{21}\left(\Omega\right)}{|\Lambda_{21}\left(\xi^{ 21}_{1}\right)|}+\Theta\left(-\Omega\right)\Theta\left(\Omega^{2}-a^{2}_{1}\right)\times\right.
  	\nonumber\\
  	&&\left.\times\frac{{\cal{F}}_{34}\left(\Omega\right)}{|\Lambda_{34}\left(\xi^{ 34}_{1}\right)|}+\Theta\left(\Omega\right)\Theta\left(\Omega^{2}-a^{2}_{1}\right)\frac{{\cal{F}}_{43}\left(\Omega\right)}{|\Lambda_{43}\left(\xi^{ 43}_{1}\right)|}+\Theta\left(a(\Omega)\right)\Theta\left(\Gamma_{a}(\Omega)\right)\Theta\left(\tilde{\Gamma}_{a}(\Omega)\right)\frac{{\cal{F}}_{23}\left(\Omega\right)}{|\Lambda_{23}\left(\xi^{ 23}_{1}\right)|}\right.
  		\nonumber\\
  		&&
  	\left.+\Theta\left(-b(\Omega)\right)\Theta\left(\Gamma_{b}(\Omega)\right)\Theta\left(\tilde{\Gamma}_{b}(\Omega)\right)\frac{{\cal{F}}_{32}\left(\Omega\right)}{|\Lambda_{32}\left(\xi^{ 32}_{1}\right)|}+\Theta\left(-a(\Omega)\right)\Theta\left(\Gamma_{a}(\Omega)\right)\Theta\left(\tilde{\Gamma}_{a}(\Omega)\right)\frac{{\cal{F}}_{14}\left(\Omega\right)}{|\Lambda_{14}\left(\xi^{ 14}_{1}\right)|}
  	\right.
  	\nonumber\\
  	&&\left.
  	+\Theta\left(b(\Omega)\right)\Theta\left(\Gamma_{b}(\Omega)\right)\Theta\left(\tilde{\Gamma}_{b}(\Omega)\right)\frac{{\cal{F}}_{41}\left(\Omega\right)}{|\Lambda_{41}\left(\xi^{41}_{1}\right)|}\right].
  	\nonumber\\
  	\label{Equation_45}
  	\end{eqnarray} 
  \end{widetext}
 Here, $\Theta(x)$ is the Heaviside's unit step function, and 
  \begin{eqnarray}
  \Gamma_{\alpha}(\Omega)=\sqrt{\alpha^{4}(\Omega)+4a^{2}_{1}b^{2}_{1}}-a^{2}_{1}-b^{2}_{1},
  \nonumber\\
  \tilde{\Gamma}_{\alpha}(\Omega)=\left(\alpha^{2}(\Omega)-a^{2}_{1}-b^{2}_{1}\right)^{2}-4a^{2}_{1}b^{2}_{1}
  \end{eqnarray}
with $\alpha=a,b$. The frequency dependent parameters $a(\Omega)$, $b(\Omega)$ and the  interaction induced parameters $a_{1}$, $b_{1}$ are defined in the following way
  \begin{eqnarray}
 && a(\Omega)=2\left(\Omega+\Delta+\gamma_1\right),
  \nonumber\\
  &&b(\Omega)=2\left(\Omega-\Delta-\gamma_1\right),
  \nonumber\\
 && a_1=V+\Delta+\gamma_1,
  \nonumber\\
  &&b_{1}=V-\Delta-\gamma_1.
  \label{Equation_46}
  \end{eqnarray}
The frequency dependent functions ${\cal{F}}_{mn}(\Omega)$ in Eq.(\ref{Equation_27}) with $m,n=1,4$ are defined as:
  \begin{widetext}
\begin{eqnarray} 
{\cal{F}}_{ mn}=\rho\left(\xi^{ mn}_{1}\right)P_{ mn}\left(\xi^{ mn}_{1}\right)\left[n_{F}\left(\mu+\Omega-\varepsilon_{m}(\xi^{ mn}_{1})\right)-n_{F}\left(\mu-\varepsilon_{m}\left(\xi^{ mn}_{1}\right)\right)\right],
\nonumber\\
+\rho\left(\xi^{ mn}_{2}\right)P_{ mn}\left(\xi^{ mn}_{2}\right)\left[ n_{F}\left(\mu+\Omega-\varepsilon_{m}(\xi^{ mn}_{2})\right)-n_{F}\left(\mu-\varepsilon_{m}\left(\xi^{ mn}_{2}\right)\right)\right],
\nonumber\\
\label{Equation_47}
\end{eqnarray}
  \end{widetext} 
where the index permutation function $P_{ mn}(x)$, in Eq.(\ref{Equation_47}) is given previously in Eq.(\ref{Equation_44}), in the precedent Section. The arguments $\xi^{ mn}_{1,2}$, in Eq.(\ref{Equation_47}), are the solutions of the equation $\Omega+\varepsilon_{n}-\varepsilon_{m}=0$. These solutions, together with the functions $\Lambda_{ mn}(\xi^{ mn}_{1})$ in Eq.(\ref{Equation_46}), are different for different optical transitions in the system, and for each transition $m\rightarrow n$ they should be separately specified. We have
\begin{eqnarray}
\xi^{ 12}_{1,2}=\xi^{ 21}_{1,2}=\pm\frac{1}{2\gamma_0}\sqrt{\Omega^{2}-b^{2}_{1}},
\nonumber\\
\xi^{ 34}_{1,2}=\xi^{ 43}_{1,2}=\pm\frac{1}{2\gamma_0}\sqrt{\Omega^{2}-a^{2}_{1}},
\nonumber\\
\xi^{ 23}_{1,2}=\pm\frac{1}{4|a(\Omega)|\gamma_0}\sqrt{(a^{2}(\Omega)-a^{2}_{1}-b^{2}_{1})^{2}-4a^{2}_{1}b^{2}_{1}},
\nonumber\\
\xi^{ 32}_{1,2}=\pm\frac{1}{4|b(\Omega)|\gamma_0}\sqrt{(b^{2}(\Omega)-a^{2}_{1}-b^{2}_{1})^{2}-4a^{2}_{1}b^{2}_{1}},
\nonumber\\
\xi^{ 14}_{1,2}=\pm\frac{1}{4|a(\Omega)|\gamma_0}\sqrt{(a^{2}(\Omega)-a^{2}_{1}-b^{2}_{1})^{2}-4a^{2}_{1}b^{2}_{1}},
\nonumber\\
\xi^{ 41}_{1,2}=\xi^{ 32}_{1,2}.
\nonumber\\
\label{Equation_48}
\end{eqnarray}  
Next, the functions $\Lambda_{ mn}(x)$ in Eq.(\ref{Equation_45}) with $m,n=1,4$ are expressed with the help of the parameters $a_{1}$ and $b_{1}$, given in Eq.(\ref{Equation_28}). We obtain
\begin{eqnarray} 
&&\Lambda_{12}(x)=-\Lambda_{21}(x)=\frac{4\gamma^{2}_{0}x}{\sqrt{b^{2}_{1}+4x^{2}\gamma^{2}_{0}}},
\nonumber\\
&&\Lambda_{34}(x)=-\Lambda_{43}(x)=\frac{4\gamma^{2}_{0}x}{\sqrt{a^{2}_{1}+4x^{2}\gamma^{2}_{0}}},
\nonumber\\
&&\Lambda_{23}(x)=-\Lambda_{32}(x)=\frac{2\gamma^{2}_{0}x}{\sqrt{a^{2}_{1}+4x^{2}\gamma^{2}_{0}}}+\frac{2\gamma^{2}_{0}x}{\sqrt{b^{2}_{1}+4x^{2}\gamma^{2}_{0}}},
\nonumber\\
&&\Lambda_{14}(x)=-\Lambda_{41}(x)=\Lambda_{23}(x).
\nonumber\\
\label{Equation_49}
\end{eqnarray}
We see here that all functions $\Lambda_{ mn}(\xi^{ mn}_{1})$ entering in the equation of the optical conductivity in Eq.(\ref{Equation_45}), depend on frequency $\Omega$ via the solutions $\xi^{ mn}_{1,2}$, given in Eq.(\ref{Equation_48}).

We have considered here the pure BLG system. The inclusion of the effects of the scattering by dilute charged impurities could be done in the self-consistent Born approximation \cite{cite_57, cite_58, cite_59}. In this case, the effect of the finite external bias voltage should be included directly and the dependence of optical conductivity on the impurity concentration and bias voltage should be considered. The universal, disorder independent, electrical conductivity at low temperatures will be relevant in this case, leading to the localized states near the Fermi level in the electronic density of states. According Born approximation in the scattering theory \cite{cite_60}, the electronic self-energy matrix $\Sigma(i\nu_{n})$ of disordered system in the presence of finite but small density impurity atoms is expressed in terms of the local Green function of clean system ${\mathcal{G}}({\bf{k}},i\nu{n})$. Then the perturbative expansion for the Green's function $\bar{{\mathcal{G}}}({\bf{k}},i\nu_{n})$ of disordered system could be obtained via the Dyson equation given as 
\begin{eqnarray} 
\bar{{\mathcal{G}}}({\bf{k}},i\nu_{n})=\frac{1}{{\mathcal{G}}^{-1}({\bf{k}},i\nu_{n})-\Sigma(i\nu_{n})}.
\end{eqnarray}
The scattering rate generated by the effect of the disorder has the energy of the order of the Fermi energy. Therefore the inclusion of the effects of disorder will have a strong impact on the optical properties in the BLG, particularly on the optical conductivity. The study of the effects of the disorder is extremely important for the technological applications of the BLG because it is almost sensitive to the unavoidable disorder created by the substrate on which it is deposited, adatoms, ionized impurities, dopping, etc.
The inclusion of the effect of disorder in our problem of the optical conductivity in the presence of the excitonic pairing could be done within the bilayer Hubbard model considered here. Especially, the large bandwidth of the Coulomb interaction potentials $U$ and $V$ is possible to consider within the present approach which will correspond to different screening regimes of the external potential by the impurity atoms in the BLG. Unfortunately, this subject it is out of the scope of the present paper. 
%
\section{\label{sec:Section_4} The numerical results}
\begin{figure}
	\begin{center}
		\includegraphics[scale=0.3]{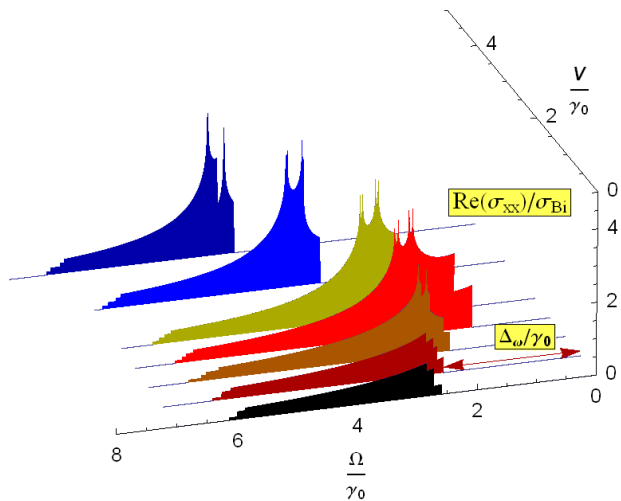}
		\caption{\label{fig:Fig_2}(Color online) The real part of the normalized longitudinal conductivity function in the bilayer graphene, given in Eq.(\ref{Equation_45}), for different values of the Coulomb interaction parameter $V$ (the plots in black, dark-red, dark-orange, red, dark-yellow, blue and dark-blue correspond to the values $V=0$, $V=0.5\gamma_0$, $V=\gamma_0$, $V_{c}=1.49\gamma_0$, $V=2\gamma_0$, $V=3\gamma_{0}$ and $V=4\gamma_0$).}.
	\end{center}
\end{figure}

%
In Fig.~\ref{fig:Fig_2}, we have presented the numerical results for the real part of the longitudinal optical conductivity function, normalized to the dc conductivity in the bilayer graphene $\sigma_{ Bi}=e^{2}/2\hbar$. Different values of the interlayer Coulomb interaction parameter are considered (from zero up to intermediate $V_{c}=1.49\gamma_0=4.47$ eV and higher values) in Fig.~\ref{fig:Fig_2} and the zero temperature limit is set for the presented plots. The intralayer Coulomb interaction parameter is fixed at the value $U=2\gamma_0=6$ eV, and the interlayer hopping amplitude is $\gamma_1=0.128\gamma_0=0.384$ eV.
%
%
%
\begin{table*}[t]
	\centering
	\begin{tabular}{{c|c|c|c|c|c|c|c|c|c}}
		$V$ & 0 & 0.5$\gamma_0$ (1.5 eV)& $\gamma_0$ (3 eV)& 1.2$\gamma_0$ (3.6 eV) & 1.49$\gamma_0$(4.47 eV) & 1.8$\gamma_0$(5.4 eV) & 2$\gamma_0$ (6 eV) & 3$\gamma_0$ (9
		eV) & 4$\gamma_0$(12 eV)\\
		\hline
		\\
		$\mu$ (eV) & -5.589  & -5.94  & -6.3  & -6.21 & -1.47 & -1.26  & -1.17 & -1.02  &-1.035 \\
		$\Delta_{\omega}$ (eV) & 7.77 & 6.93 & 5.865 & 4.917  & 4.011 & 5.4 & 6.174 & 9.483 &
		12.48\\
		$\varepsilon_{F}$ (eV) & -4.089 & 3.69 & -3.3 & -2.91 & 2.265 & 2.94 & 3.33 & 4.98 &  
		6.465\\
	\end{tabular}
	\caption{The exact solutions of the chemical potential $\mu$, optical gap $\Delta_{\omega}$ and Fermi energy $\varepsilon_{F}$.}
	\label{tab:a}
\end{table*}
%
\begin{figure}
	\begin{center}
		\includegraphics[scale=0.65]{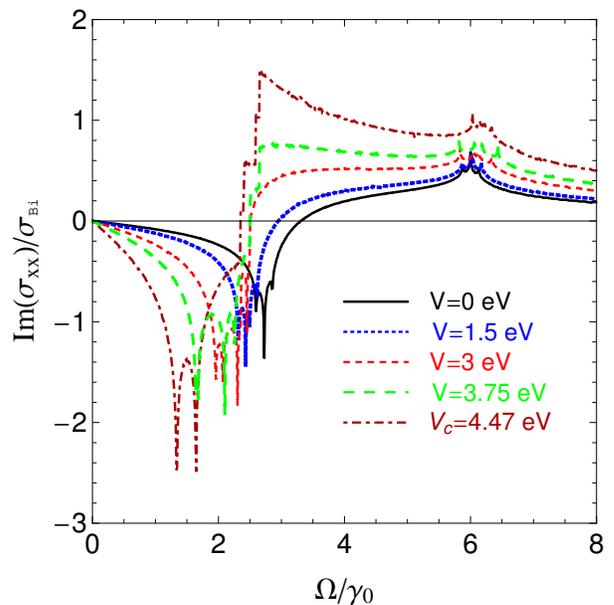}
		\caption{\label{fig:Fig_3}(Color online) The imaginary part of the normalized longitudinal conductivity function in the bilayer graphene, given in Eq.(\ref{Equation_50}). Different values of the interlayer Coulomb interaction parameter are considered (from zero up to $V_c$).}.
	\end{center}
\end{figure} 
%
In table \ref{tab:a}, we give the obtained values of the chemical potential \cite{cite_52}, the threshold frequency values (the optical gaps) for the real part of the conductivity function, and the Fermi energy in the BLG corresponding to the interlayer interaction parameters, considered in Fig. ~\ref{fig:Fig_2}.
From the results, given in Fig.~\ref{fig:Fig_2}, and table \ref{tab:a}, we can see the general behavior of the real part of the optical conductivity function, as a function of the chemical potential $\mu$ and Fermi level $\varepsilon_{F}$ in the bilayer graphene. First of all, let's mention that the BLG system is automatically extrinsic in our case (this is the case when $\varepsilon_{F}\neq 0$ because of the finite number of the electron-hole pairs), in difference with the discussion in Ref.\onlinecite{cite_40}, where both intrinsic and extrinsic cases have been considered equally, and the optical gap has been attributed to the first large peak in the intrinsic optical conductivity spectrum. Particularly, for all values of the interaction parameter $V$, we observe the presence of a very large optical gap in the real part of the conductivity spectrum in Fig.~\ref{fig:Fig_2}. Considering the case of the noninteracting layers, i.e., when $V=0$, we see that the highest peak in optical conductivity spectrum (see the deepest black plot in Fig.~\ref{fig:Fig_2}) is situated at the value $\Omega=2.856\gamma_0$, or at the high-energy resonant position at $\Omega=8.57$ eV, which is very close to the excitonic resonance at $8.3$ eV obtained in recent ab-initio many-body calculations of the high energy optical absorption in graphene \cite{cite_46}. Moreover, when augmenting the interaction parameter from zero up to the critical value $V_{c}=1.49\gamma_0=4.47$ eV the optical gap is decreasing, while the optical conductivity is largely increased, and the peaks positions become more apparent. The value $V_{c}=1.49\gamma_0$, at which the optical gap attains its minimum $\Delta_{\omega_{0}}=1.337\gamma_0=4.011 $ eV, (corresponding to the photon wavelength $\lambda=309.15$ nm, in the near UV range of the spectrum), is the threshold value of $V$, above which the chemical potential and the Fermi level jump to their upper bound solutions (for details, see in Refs.\onlinecite{cite_11}, and \onlinecite{cite_52}) and the critical value $V _c=1.49\gamma_0$ plays the role of the new CNP in the interacting BLG, as it was explained in Ref.\onlinecite{cite_11}. Let's mention also that the value of $\Delta_{\omega_{0}}$ at $V_c$ coincides exactly with the value of the hybridization gap $\Delta_{H}$ in the BLG, obtained in Ref.\onlinecite{cite_52}). 

Thus, we observe that for $V<V_c$ the interaction effect acts to redshift the conductivity peaks in the BLG. For $\Omega<\Delta_{\omega_{0}}$, we have $\Re\sigma_{ xx}(\Omega)/\sigma_{ Bi}=0$. Furthermore, starting from the CNP value of $V$, the optical gap is increasing (see in Fig.~\ref{fig:Fig_2}) and becomes very large at $V=4\gamma_0=12$ eV, of the order of $\Delta_{\omega}=12.48$ eV. On the other hand, there are no remarkable changes in the conductivity peaks amplitudes, in this case, corresponding to the different optical transitions in the bilayer graphene. The other important feature in the strong interaction regime is related to the blue- shift effect of the conductivity peaks above the CNP value of $V$, i.e. when $V>V_{c}$. Thus, the general conclusion is that when the chemical potential and the Fermi energy in the BLG pass to their upper bound solutions as a function of the interlayer interaction parameter (see about in Refs.\onlinecite{cite_11} and \onlinecite{cite_12} and also in Table \ref{tab:a}), the excitonic pairing causes to change the red-shift effect of the conductivity peaks into the blue-shifted ones. 

From zero up to critical value of the interlayer interaction parameter $V\in[0;V_c]$, the optically active photon frequency region is given by the frequency interval $\Omega\in[4.011;19]$ eV, and corresponds to the photon's wavelengths $\lambda\in[65.26;309.15]$ nm, thus situating in the range, which starts from the extreme ultraviolet part of the light spectrum and increases up to near UV region. We observe also that inside this region of the interaction parameter, there is no conductivity spectrum displacement in the far UV side of the spectrum, and the observable changes occur only in the near UV part of the photon energies, governing the changes of the optical gap $\Delta_{\omega}$. Contrary, for the strong interlayer interaction regime, the considerable changes occur on both sides of the spectrum. For the large interaction values, considered in Fig.~\ref{fig:Fig_2}, the conductivity spectrum is squeezing when increasing the interaction parameter $V$, and the near UV part of the spectrum gets displaced into the smallest wavelengths sides, i.e., along with the UV-c part of the spectrum. For the very strong interaction $V=4\gamma_0=12$ eV, the corresponding active optical frequencies are given by the energy interval $\Omega\in [12.47;22]$ eV and the corresponding wavelengths are given in $\lambda\in[56.17; 99.43]$ nm. Thus, in this case, the strong excitonic effects are present in the system and, experimentally, it would be quite difficult to observe them in this limit because of the very narrow region of the permitted photon's wavelengths.    

The imaginary part of the optical conductivity function $\sigma_{ xx}(\Omega)$ can be easily calculated using the Kramers-Kronig formula, which relates the real and imaginary parts of the complex optical conductivity function. It is given as 
\begin{eqnarray}
{\Im\sigma_{ xx}}(\Omega)=-\frac{2\Omega}{\pi}\int^{\infty}_{0}d\Omega'\frac{\Re\sigma_{ xx}(\Omega')}{\Omega'^{2}-\Omega^{2}},
\label{Equation_50}
\end{eqnarray}  
where a special attention should be paid to the singularity points $\Omega'=\pm\Omega$ when performing the numerical integration.
The plots of the imaginary part of the conductivity function, normalized to the dc conductivity of the BLG $\sigma_{\rm Bi}$ and calculated with help of the relation in Eq.(\ref{Equation_50}) are given in Fig.~\ref{fig:Fig_3}. The function $\Im\sigma_{ xx}(\Omega)/\sigma_{ Bi}$ has been evaluated for different values of the interlayer interaction parameter (from zero up to the critical value $V_{c}=1.49\gamma_0$) and is presented as a function of the normalized frequency $\Omega/\gamma_0$. We see that the positions of the negative conductivity peaks (see in Fig.~\ref{fig:Fig_3}) are displacing to the lower-$\Omega$ regions when augmenting the interaction parameter $V$ up to critical value $V_{c}=1.49\gamma_0$ which is related to the critical value of the lower bound solution of the chemical potential at the CNP: $\mu_{c}=-0.499\gamma_0=-1.498$ eV (see in Ref.\onlinecite{cite_11}). 

The amplitudes of peaks are increasing when increasing the interaction parameter from $0$ up to $V_{c}$. The multiple peak structure in the imaginary conductivity spectrum (see the peaks structures in the curves corresponding to the negative values of $\Im\sigma_{ xx}(\Omega)/\sigma_{ Bi}$) is the artifact of the strong excitonic effects in the bilayer graphene. The important observation that could be gained from the results in Figs.~\ref{fig:Fig_2} and ~\ref{fig:Fig_3} and from the values given in table \ref{tab:a} is that the optical gap fits well with the formula 
\begin{eqnarray}
\Delta_{\omega}=2\varepsilon_{F}-\gamma_1.
\label{Equation_51}
\end{eqnarray}  

The shift by $\gamma_1$ in Eq.(\ref{Equation_51}) is due to the interlayer hopping. This result corresponds approximatively to the interband absorption threshold of the optical conductivity in the statically screened Thomas-Fermi regime, which has been well discussed in Ref.\onlinecite{cite_23}, where the Keldysh transport formalism has been used to consider the influence of the band renormalization and excitonic electron-electron interaction effects on the optical conductivity in doped bilayer graphene. The Mahan's exciton bound state vanishes in our case due to the strong interaction effects, present in the BLG. We considered here the suspended bilayer graphene where the interaction effects have a spectacular impact on the conductivity spectrum because the dielectric environment portion of the screening is absent.

\begin{figure}
	\begin{center}
		\includegraphics[scale=0.65]{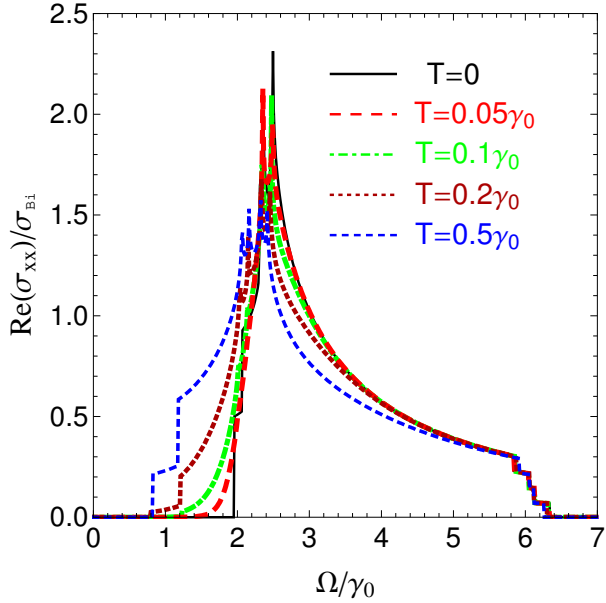}
		\caption{\label{fig:Fig_4}(Color online) The real part of the optical conductivity function, given in Eq.(\ref{Equation_45}) for different values of normalized temperature $T/\gamma_0$. The interlayer Coulomb interaction parameter is set at $V=\gamma_0=3$ eV.}.
	\end{center}
\end{figure}

\begin{figure}
	\begin{center}
		\includegraphics[scale=0.65]{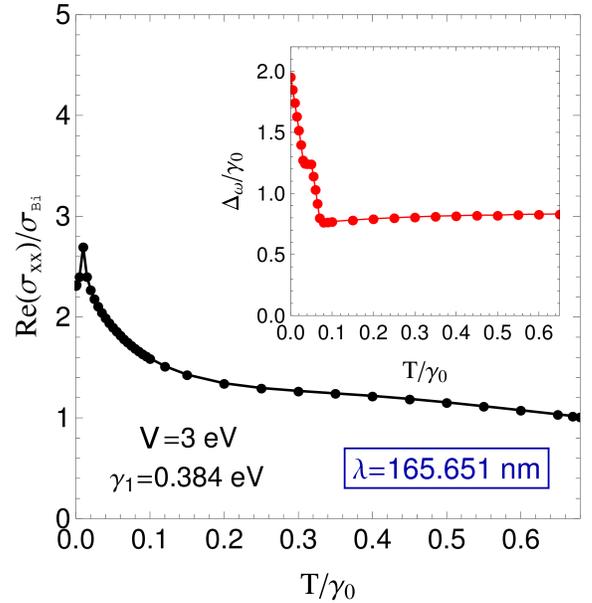}
		\caption{\label{fig:Fig_5}(Color online) Temperature dependence of the real part of the optical conductivity function, given in Eq.(\ref{Equation_45}). The interlayer interaction parameter is set at the value $V=3$ eV and the incident photon wavelength is fixed at $\lambda=165.651$ nm. Inset: temperature dependence of the optical gap $\Delta_{\omega}$ is shown for the same value of the interlayer coupling parameter.}.
	\end{center}
\end{figure}

In Fig.~\ref{fig:Fig_4}, we have presented the optical conductivity function for different values of temperature. The interlayer interaction parameter is fixed at the value $V=\gamma_0$, corresponding to the plot in orange, in Fig.~\ref{fig:Fig_2} and the curve in red in Fig.~\ref{fig:Fig_3}. We observe that when increasing the temperature, the far-UV part of the spectrum is unchanged, while the near UV-c side of the spectrum (simultaneously with the optical gap) is considerably modified and spread over the visible region of the photon's spectrum, i.e., for example, at the very high temperature $T=0.5\gamma_0$ the photon wavelength corresponding to the optical gap $\Delta_{\omega}=0.822\gamma_0=2.46$ eV, is of order $\lambda=1240/\Delta_{\omega}=502.8$ nm, situating in the visible range of the light spectrum. We see that the optical gap and the amplitudes of the conductivity peaks are decreasing when increasing the temperature. In Fig.~\ref{fig:Fig_5}, we give the explicit temperature dependence of the real part of the optical conductivity function for the fixed interaction parameter $V=\gamma_0=3$ eV and for the incident photon wavelength of the order of $\lambda=165.651$ nm. For that given wavelength, the optical conductivity function attains its dc limit $\sigma_{ Bi}$ only at the very high temperatures, of the order of $T\sim 0.7\gamma_0$.
The temperature dependence of the optical gap $\Delta_{\omega}$ (within the same temperature range)
is shown in the inset, in Fig.~\ref{fig:Fig_5}, for the case $V=3$ eV. We see that for the temperatures $T>0.07\gamma_0$, the optical gap varies very slowly with temperature.
%
%
%
\section{\label{sec:Section_5} Concluding remarks}
%
We have calculated the optical conductivity function in the interacting bilayer graphene system, exposed in the external longitudinal electric field and in the presence of the excitonic pairing interaction in the system. We have used the exact four-band model of the BLG system, by avoiding low-energy approximation in the band dispersion spectrum.  
We have shown that for a fixed value of the interlayer hopping amplitude $\gamma_1$, the excitonic pairing interaction affects considerably the optical conductivity spectrum in the bilayer graphene, which becomes more pronounced when increasing the interlayer interaction parameter $V$. We have shown that the inclusion of the excitonic effects in the system leads to the total suppression of the low-frequency Drude-contribution, related to the intraband optical transitions in the system. We have calculated the optical conductivity function for different values of the interlayer interaction parameter, and we have shown the existence of the optical gap in the extrinsic case of the bilayer graphene. The spectacular features in the conductivity spectrum are related to the behavior of the chemical potential and Fermi energy in the BLG when modifying the interlayer coupling parameter $V$. 

At the CNP, corresponding to the critical interlayer coupling strength $V_{c}$, the Fermi energy passes into its upper bound solution through CNP, and the optical conductivity spectra get switched inversely, turning from the redshifted lines into the blueshifted ones. This effect dominates along the whole optical spectrum, and we suggest that it should affect considerably also on the other optical properties in the bilayer graphene such as the refractive index, the optical absorption spectra, and the optical reflectance. The general rule, obtained in the present paper, is that the optical conductivity spectrum and the optical gap parameter are strongly governed by the position of the Fermi energy in the interacting bilayer graphene.

The obtained results could represent a considerable interest when studying experimentally the excitonic effects in the optical conductivity spectrum in the strongly interacting bilayer graphene. Furthermore, the results obtained here would represent as the reference for the calculation of the other optical properties in the system such as the optical absorption spectra, reflectivity, and the electron energy loss spectra. The results obtained in the present study could spread a new insight into the possible optoelectronic and nano-optical applications of the bilayer graphene.
%
\section{\label{sec:Section_6} Author contribution statement}
%
All authors contributed equally to the paper.
\appendix

\section{\label{sec:Section_7} The current-current operator}
%
The paramagnetic current operator $j_{i}(\mathbf{r})$, given in Eq.(\ref{Equation_31}), describes the in-plane electric current density, driven in the BLG layers by the external electric field component. We present here the calculation of the real part of the ac conductivity $\sigma_{ij}(\Omega)$ in the BLG with the help of the retarded polarization function $\Pi^{R}_{ij}(\Omega)$ and by using the standard Kubo-Green-Matsubara formalism \cite{cite_47} 
\begin{eqnarray}
\Re{\sigma_{ij}(\Omega)}=\frac{\Im{\Pi^{R}_{ij}(\Omega)}}{\Omega},
\label{Equation_A_1}
\end{eqnarray}
where 
\begin{eqnarray}
\Pi^{R}_{ij}(\Omega)=\Pi^{R}_{ij}(i\omega_m\rightarrow \Omega+i\eta^{+})
\label{Equation_A_2}
\end{eqnarray}
is the real frequency retarded function, which is obtained after the analytical continuation of the bosonic Matsubara frequencies $i\omega_m$ in to the upper half of the real axis in the complex plane, i.e., $i\omega_m\rightarrow \Omega+i\eta^{+}$ with $\eta^{+}$ being the infinitesimal positive constant, and the bosonic Matsubara frequencies $\omega_m$ are give as usual by $\omega_m=2\pi{m}/\beta$, where $m=0,\pm1,\pm2,...$, and $\beta=1/{k_{B}T}$. The current-current correlation function, in turn, is given in the Kubo-Green-Matsubara formalism \cite{cite_47} as
\begin{eqnarray}
\Pi^{R}_{ij}(i\omega_m)=-\int^{\beta}_{0}d\tau e^{i\omega_m\tau}\left\langle T_{\tau}j_{i}(\tau)j_{j}(0)\right\rangle,
\label{Equation_A_3}
\end{eqnarray}
where $\tau$ is the imaginary time, and $T_{\tau}$ is the chronological time ordering operator \cite{cite_47, cite_50}. The response function $\Pi^{R}_{ ij}(i\omega_m)$ in Eq.(\ref{Equation_A_3}) corresponds to the particle-hole bubble, in terms of Matsubara Green's functions. 
The current operator $j_{i}(\tau)$, in Eq.(\ref{Equation_A_3}), could be obtained after summing over the lattice sites in the operator given in Eq.(\ref{Equation_31}). We have
\begin{eqnarray}
j_{i}(\tau)=\sum_{\mathbf{r}}j_{i}({\mathbf{r}},\tau).
\label{Equation_A_4}
\end{eqnarray}
Here, we have attributed the imaginary time variables to the creation and annihilation operators, in Eq.(\ref{Equation_31}). After transforming the r.h.s. in Eq.(\ref{Equation_4}) into the Fourier $\mathbf{k}$-space, and after summing over the lattice sites positions, we get the following expression for the total current density operator
\begin{eqnarray}
j_{i}(\tau)=\frac{1}{N}\sum_{\mathbf{k}}\left[{v}^{\ast}_{\mathbf{k}i}a^{\dag}_{1\mathbf{k}}(\tau)b_{1\mathbf{k}}(\tau)+{v}_{\mathbf{k}i}b^{\dag}_{1\mathbf{k}}(\tau)a_{1\mathbf{k}}(\tau)\right.
\nonumber\\
\left.+{v}^{\ast}_{\mathbf{k}i}a^{\dag}_{2\mathbf{k}}(\tau)b_{2\mathbf{k}}(\tau)+{v}_{\mathbf{k}i}b^{\dag}_{2\mathbf{k}}(\tau)a_{2\mathbf{k}}(\tau)\right],
\label{Equation_A_5}
\end{eqnarray}
where
\begin{eqnarray}
{v}_{\mathbf{k}i}=-\frac{i\gamma_0}{\hbar}\sum_{\bm{\mathit{\delta}}}\bm{\mathit{\delta}}_{i}e^{i{\mathbf{k}}\bm{\mathit{\delta}}}
\label{Equation_A_6}
\end{eqnarray}
is the electron velocity operator in the individual graphene sheet. There exist several theoretical approaches for treating the BLG in the presence of the external fields. Usually one considers the applied external bias voltage \cite{cite_20, cite_24}, which makes the considerable potential difference between the layers in the BLG. Moreover, it leads to the charge imbalance between the layers. The electron Coulomb interaction between the layers was not considered is such approaches \cite{cite_20, cite_24} and the tight-binding description is discussed only.
The authors found a very large asymmetry gap, induced in the quasiparticle excitation spectrum, which leads to the charge-fluctuations in the system. In another treatment \cite{cite_11, cite_12}, the intralayer and interlayer Coulomb interactions have been included properly in the Hamiltonian of the bilayer graphene, given in Eq.(\ref{Equation_1}), here. 

We follow here the second routine, described above, in order to calculate the ac conductivity in the interacting bilayer graphene with the presence of the excitonic pairing interaction. 
We perform the full bandwidth 4-band calculation scheme, which goes beyond the usually known Dirac-cone approximation \cite{cite_3}. We will concentrate here on the calculation of the four-point fermionic correlation functions that appear after putting Eq.(\ref{Equation_A_5}) in the expression of the polarization function, given in Eq.(\ref{Equation_A_3}). We consider here only the $x$ component of the vector potential, thus the current density operator is given in the same direction $j_{x}(\tau)$ as the electric field component $E_x$, which oscillates parallel to the graphene's layers. Next, after performing the Wick averaging procedure \cite{cite_51}, we get a very compact expression for the current-current correlation function 
\begin{eqnarray}
&& \left\langle
T_{\tau}j_{x}(\tau)j_{x}(0)\right\rangle=
\nonumber\\ &&-{2e^{2}}\sum_{\mathbf{k}\mathbf{k}'}\left[         
v^{\ast}_{\mathbf{k}x}v_{\mathbf{k}'x}{\cal{G}}_{ aa}(\mathbf{k}'0,\mathbf{k}\tau){\cal{G}}_{ bb}(\mathbf{k}\tau,\mathbf{k}'0)\right.
\nonumber\\
&&\left.+v_{\mathbf{k}x}v^{\ast}_{\mathbf{k}'x}{\cal{G}}_{ bb}(\mathbf{k}'0,\mathbf{k}\tau){\cal{G}}_{ aa}(\mathbf{k}\tau,\mathbf{k}'0)\right.
\nonumber\\
&& \left. +v^{\ast}_{\mathbf{k}x}v_{\mathbf{k}'x}{\cal{G}}_{ \tilde{a}\tilde{a}}(\mathbf{k}'0,\mathbf{k}\tau){\cal{G}}_{ \tilde{b}\tilde{b}}(\mathbf{k}\tau,\mathbf{k}'0)\right.
\nonumber\\
&&\left.+v_{\mathbf{k}x}v^{\ast}_{\mathbf{k}'x}{\cal{G}}_{ \tilde{b}\tilde{b}}(\mathbf{k}'0,\mathbf{k}\tau){\cal{G}}_{ \tilde{a}\tilde{a}}(\mathbf{k}\tau,\mathbf{k}'0)\right].
\label{Equation_A_7}
\end{eqnarray}  
The single-particle Green's functions ${\cal{G}}_{ {a}{a}}$ and ${\cal{G}}_{ {b}{b}}$ are non zero only if $\mathbf{k}=\mathbf{k}'$, therefore, ${\cal{G}}_{ {a}{a}}(\mathbf{k}'0,\mathbf{k}\tau)=\delta_{\mathbf{k}\mathbf{k}'}{\cal{G}}_{ {a}{a}}(\mathbf{k}0,\mathbf{k}\tau)$. This follows from the symmetry of the action, given in Eq.(\ref{Equation_6}), in the Section \ref{sec:Section_2_1} of the present paper.
%

%

\begin{thebibliography}{10}
%
\bibitem{cite_1} Liu, J.‐M. Principles of Photonics. Cambridge, United Kingdom: Cambridge University
Press; (2016). 260 p. DOI: 10.1017/CBO9781316687109.
\bibitem{cite_2} K. Novoselov, A. K. Geim, S. V. Morozov, D. Jiang, M. I.
Katsnelson, I. V. Grigorieva, S. V. Dubonos, and A. A. Firsov, Nature London 438, 197 (2005).
\bibitem{cite_3} A. H. Castro Neto, F. Guinea, N. M. R. Peres, K. S. Novoselov, A. K. Geim, Rev. Mod. Phys. 81 109 (2009).
\bibitem{cite_4} E. V. Castro, et al. Phys. Rev. Lett 99 216802 (2007).
\bibitem{cite_5} J. Nilsson, A. H. Castro Neto, F. Guinea and N. M. R. Peres Phys. Rev. B 76 165416 (2007).
\bibitem{cite_6} H. Leal, D. V. Khveshchenko, Nucl. Phys. B 687, 323 (2004).
\bibitem{cite_7} I. L. Aleiner, D.E. Kharzeev, A. M. Tsvelik, Phys. Rev. B 76, 195415 (2007).
\bibitem{cite_8} D. V. Khveshchenko, Phys. Rev. Lett. 87, 206401 (2001).
\bibitem{cite_9} D. V. Khveshchenko, J. Phys. Condens. Matter 21, 075303 (2009).
\bibitem{cite_10} E. V. Gorbar, V. P. Gusynin, V. A. Miransky, I. A. Shovkovy, Phys. Lett. A 313, 472 (2003).
\bibitem{cite_11} V. Apinyan, T. K. Kope\'{c}, Phys. Scr. 91, 095801 (2016).
\bibitem{cite_12} V. Apinyan, T. K. and Kope\'{c}, Eur. Phys. J. B 90 130 (2017).
\bibitem{cite_13} F. Wang, Y. Zhang, C. Tian, C. Girit, A. Zettl, M. Crommie, and Y. R. Shen, Science 320, 206 (2008).
\bibitem{cite_14} B. Huard et al., Phys. Rev. Lett. 98, 236803 (2007).
\bibitem{cite_15} J. R. Williams, L. DiCarlo, C. M. Marcus, Science 317, 638 (2007).
\bibitem{cite_16} D. S. L. Abergel and Vladimir I. Fal'ko,  Phys. Rev. B 75, 155430 (2007).
\bibitem{cite_17} E. V. Gorbar, V. P. Gusynin, A. B. Kuzmenko, and S. G. Sharapov, Phys. Rev. B 86, 075414 (2012).
\bibitem{cite_18} J. Nilsson, A. H. Castro Neto, F. Guinea, and N. M. R. Peres, Phys. Rev. Lett. 97, 266801 (2006). 
\bibitem{cite_19} J. Cserti, Phys. Rev. B 75, 033405 (2007). 
\bibitem{cite_20} E. J. Nicol and J. P. Carbotte, Phys. Rev. B 77, 155409 (2008).
\bibitem{cite_21} E. McCann, D. S. Abergel, and V. I. Fal’ko, Solid State Commun. 143, 110 (2007). 
\bibitem{cite_22} C. H. Yang, Z. M. Ao,  X. F. Wei and J. J. Jiang, Physica B 457 92 (2015).
\bibitem{cite_23} Wang-Kong Tse and A. H. MacDonald,  Phys. Rev. B 80, 195418 (2009).
\bibitem{cite_24} H Rezania and M Yarmohammadi, Indian J Phys, 90(7), 811 (2016).
\bibitem{cite_25} Palash Nath, D. Sanyal, Debnarayan Jana, Current Applied Physics, 15 691e697 (2015).
\bibitem{cite_26} C.H. Yang, Y. Y. Chen, J. J. Jiang, Z. M. Ao, Solid State Communications 227 23–27 (2016).
\bibitem{cite_27} D. S. L. Abergel, A. Russell, and Vladimir I. Fal’ko, Applied Physics Letters 91, 063125 (2007).  
\bibitem{cite_28} M. Koshino, New Journal of Physics 11 095010 (2009).
\bibitem{cite_29} L.A. Falkovsky Pis’ma v ZhETF, 97 (7), 496-505 (2013) [JETP Lett., 97(7), 429-438 (2013)].
\bibitem{cite_30} L.A. Falkovsky, Jetp, 110(2), 319-324 (2010).
\bibitem{cite_31} L. M. Zhang, Z. Q. Li, D. N. Basov, M. M. Fogler, Z. Hao and M. C. Martin Phys. Rev. B 78 235408 (2008).
\bibitem{cite_32} You-Chia Chang, Chang-Hua Liu, Che-Hung Liu, Zhaohui Zhong and Theodore B. Norris, Applied Physics Letters, 104, 261909 (2014).
\bibitem{cite_33} M. Bruna and S. Borini, Applied Physics Letters, 94, 031901 (2009).
\bibitem{cite_34} Yingying Wang, Zhenhua Ni, Lei Liu, Yanhong Liu, Chunxiao Cong, Ting Yu, Xiaojun Wang, Dezhen Shen and Zexiang Shen, ACS Nano, 4 (7), pp 4074–4080 (2010).
\bibitem{cite_35} A. B. Kuzmenko, I. Crassee, D. van der Marel, P. Blake, and K. S. Novoselov, Phys. Rev. B 80, 165406 (2009).
\bibitem{cite_36} Z. Q. Li, E. A. Henriksen, Z. Jiang, Z. Hao, M. C. Martin, P. Kim, H. L. Stormer, and D. N. Basov, Phys. Rev. Lett. 102, 037403 (2009).
\bibitem{cite_37} K. F. Mak, C. H. Lui, J. Shan, and T. F. Heinz, Phys. Rev. Lett. 102,
256405 (2009).
\bibitem{cite_38} D. S. L. Abergel, A. Russell, and V. I. Falko, Appl. Phys. Lett. 91, 063125 (2007).
\bibitem{cite_39} T. Stauber, N. M. R. Peres, and A. K. Geim, Phys. Rev. B 78, 085432 (2008).
\bibitem{cite_40} D. S. L. Abergel, Hongki Min, E. H. Hwang, and S. Das Sarma, Phys. Rev. B 85, 045411 (2012).
\bibitem{cite_41} L. Yang, C.-H. Park, J. Deslippe, and S. G. Louie, Phys. Rev. Lett. 103, 186802 (2009).
\bibitem{cite_42} V. G. Kravets, A. N. Grigorenko, R. R. Nair, P. Blake, S. Anissimova, K. S. Novoselov, and A. K. Geim, Phys. Rev. B
81, 155413 (2010).
\bibitem{cite_43} G. Yu, J. Gao, J. C. Hummelen, F. Wudl, and A. J. Heeger, Science 270, 1789 (1995).
\bibitem{cite_44} A. Goetzberger, C. Hebling, and H. Schock, Materials Science and Engineering 40, 1 (2003).
\bibitem{cite_45} Li Yang, Phys. Rev. B 83, 085405 (2011). 
\bibitem{cite_46} P. E. Trevisanutto, M. Holzmann, M. C\^{o}t\'e, and V. Olevano, Phys. Rev. B 81, 121405 (2010). 
\bibitem{cite_47} G. D. Mahan, Many-Particle Physics, 3rd ed. (Kluwer Academic/Plenum, New York, 2000).
\bibitem{cite_48} A. J. Millis, in Strong Interactions in Low Dimensions, edited by D. Baeriswyl and L. De Giorgi (Kluver Academic, Berlin, 2003).
\bibitem{cite_49} Wannier, G., Rev. Mod. Phys. 34, 645 (1962).
\bibitem{cite_50}J. W. Negele and H. Orland, Quantum Many-Particle Systems, Addison-Wesley, Reading, MA, (1988).
\bibitem{cite_51} A. A. Abrikosov, L. P. Gorkov, I. E. Dzyaloshinski, Methods of Quantum Field Theory
in Statistical Physics, Pergamon Press, (1965).
\bibitem{cite_52} V. Apinyan, T. K. Kope\'{c}, Physica E 95, 108, (2018). 
\bibitem{cite_53} V. Apinyan, T. K. Kope\'{c}, Superlattices and Microstructures, 119, 166 (2018).
\bibitem{cite_54} J. P. Hobson and W. A. Nierenberg, Phys. Rev. 89 (1953) 662.
\bibitem{cite_55} M. Abramovitz, I. Stegun, Handbook of Mathematical Functions (Dover, New York, 1970).
\bibitem{cite_56} R. R. Nair et al., Science 320, 1308 (2008).
\bibitem{cite_57} N.M.R. peres, F. Guinea, H. Castro Neto, Phys. Rev. B 73 (2006) 125411.
\bibitem{cite_58} Nguyen Hong Shon, Tsuneya Ando, J. Phys. Soc. Jpn. 67 (1998) 2421.
\bibitem{cite_59} X.-Z. Yan, Y. Romiah, C.S. Ting, Phys. Rev. B 77 (2008) 125409. X-Z. Yan and C.
S. Ting, Phys. Rev. B 80, 155423(2009).
\bibitem{cite_60} S. Doniach, E.H. Sondheimer, (2nd ed.), Green's Functions for Solid State Physicists, Benjamin, Reading (1974).
\end{thebibliography}
\end{document}